\documentclass[10pt, aps, prx,twocolumn,nofootinbib,superscriptaddress,longbibliography]{revtex4-2}
\usepackage{hyperref}

\hypersetup{
    unicode=true, 
    colorlinks=true, 
    linkcolor=teal, 
    citecolor=magenta,
    urlcolor=blue
}

\usepackage{latexsym,amssymb,amsmath,amsthm,bm,bbold,amsfonts,amstext,graphicx,bbm,dsfont,enumerate,float,enumitem}

\usepackage{bold-extra}
\usepackage[table]{xcolor}
\usepackage[utf8]{inputenc}
\usepackage[english]{babel}
\usepackage{tikz}
\usepackage{transparent}
\usepackage[normalem]{ulem}
\usepackage{setspace}
\usepackage{textcomp}
\usepackage[most,skins]{tcolorbox}
\usepackage{tabularx}
\definecolor{light_blue}{HTML}{f0f5ff}
\usepackage{color}
\usepackage{graphicx}

\makeatletter
\def\@hangfrom#1{\setbox\@tempboxa\hbox{{#1}}%
  \hangindent 0pt%
  \noindent\box\@tempboxa}

\renewcommand\section{%
  \@startsection{section}{1}{0pt}%
    {3.5ex plus 1ex minus .2ex}%
    {2.3ex plus .2ex}%
    {\normalfont\bfseries\raggedright}%
}
\renewcommand\subsection{%
  \@startsection{subsection}{2}{0pt}%
    {3.25ex plus 1ex minus .2ex}%
    {1.5ex plus .2ex}%
    {\normalfont\bfseries\raggedright}%
}
\renewcommand\subsubsection{%
  \@startsection{subsubsection}{3}{0pt}%
    {3.25ex plus 1ex minus .2ex}%
    {1ex plus .2ex}%
    {\normalfont\itshape\raggedright}%
}
\makeatother

\newcommand{\etal}{\textit{et al.$\:$}}

\newcommand{\tr}{\mathrm{Tr}}

\def\bra#1{\langle{#1}|}
\def\ket#1{|{#1}\rangle}
\def\braket#1{\langle{#1}\rangle}
\newcommand{\ketbra}[2]{\ket{#1}\!\bra{#2}}

\newcommand{\phdagger}{{\phantom{\dagger}}}

\definecolor{boxcolor}{HTML}{e5e3fa}
\definecolor{lightgrey}{HTML}{e0e0e0}
\definecolor{basegrey}{HTML}{f1f2f2}
\definecolor{quantum_purple}{HTML}{8f05ff}
\definecolor{volt}{HTML}{f1ffcc}
\definecolor{dark_green}{HTML}{006340}
\definecolor{energy_orange}{HTML}{fc3503}
\definecolor{bright_orange}{HTML}{fff5e6}
\definecolor{energy_yellow}{HTML}{ff7300}
\definecolor{light_yellow}{HTML}{fff9e6}

\begin{document}
\sloppy 

\title{Molecular triplets and other metastable states for excitonic quantum batteries}

\author{Daniel J. Tibben}
\affiliation{School of Science, RMIT University, Melbourne, 3000, Victoria, Australia}
\affiliation{RMIT Applied Quantum Technologies Centre, RMIT University, Melbourne, 3000, Victoria, Australia}

\author{Gian Marcello Andolina}
\affiliation{EIP, UAR 3573 CNRS, Coll\`ege de France, PSL Research University, 11 Place Marcelin Berthelot, F-75321 Paris, France}

\author{Daniel E. G\'omez}
\email{daniel.gomez@rmit.edu.au}
\affiliation{School of Science, RMIT University, Melbourne, 3000, Victoria, Australia}

\author{Francesco Campaioli}
\email{francesco.campaioli@unipd.it}
\affiliation{Department of Physics, School of Science, RMIT University, Melbourne, 3000, Victoria, Australia}
\affiliation{RMIT Applied Quantum Technologies Centre, RMIT University, Melbourne, 3000, Victoria, Australia}
\affiliation{Dipartimento di Fisica e Astronomia, Universit\'a degli Studi di Padova, 35131 Padova, Italy}

\date{\today}

\begin{abstract}
Excitonic quantum batteries, based on organic fluorescent molecules embedded in optical microcavities, offer a room-temperature platform for studying collective effects in energy storage and developing applications. Recent experiments have offered evidence of superabsorption, a collective enhancement to the light absorption rate of organic molecules which leads to a scalable power density. However, they have also highlighted the challenge posed by rapid radiative decay of fluorescent molecules, which limits the energy storage lifetime. Current strategies to overcome this trade-off focus on controlling the coupling between the absorbing manifold and that used for energy storage. In this chapter, we review three implementations of this design principle: transferring energy from optically excited states to long-lived dark triplet states, generating triplet pairs and higher-spin states through singlet exciton fission, and forming charge-separated states. We discuss each mechanism from both theoretical and experimental perspectives, with particular emphasis on recent device implementations that have extended storage times by several orders of magnitude. We conclude with a cross-platform outlook on the role of metastable states across coherent and room-temperature implementations, from neutral atom arrays to masers and colour centres.
\end{abstract}

\maketitle
\makeatletter

{
\tableofcontents
}

\section{Introduction}
\label{s:introduction}

\noindent
Quantum batteries were introduced in the early 2010s by the quantum thermodynamics community as a way to explore energy storage at the quantum mechanical level~\cite{Alicki2013}. The core question driving those early investigations was whether entanglement---the nonclassical correlation at the heart of the second quantum revolution~\cite{Deutsch2020}---could offer an advantage in the tasks of charging, energy storage, and work extraction~\cite{Binder2018a}. This initial effort identified a relationship between entanglement generation and charging power in many-body quantum systems, showing that collective effects can be exploited to achieve charging speedups that scale with the number of subsystems involved~\cite{Binder2015, Campaioli2017}, i.e., batteries that charge faster as they grow in size. This counter-intuitive discovery ignited a systematic search for physically realisable models in which such  speedups could be observed~\cite{Campaioli2024, ferraro2026opportunities}.

A pivotal moment in this research area came in 2018, when Ferraro \etal{} introduced a proposal for a solid state quantum battery that could be optically charged~\cite{Ferraro2018}, based on the Dicke model of light-matter interaction, well-known for its superradiant properties~\cite{Dicke_PR1954a, GrossHaroche1982}. Ferraro \etal{} showed that a system of $N$ emitters, such as atoms or molecules, can achieve a charging power that scales as $N\sqrt{N}$ when coupled to the optical mode of a cavity already populated with $N$ or more photons, as long as the cavity volume does not scale with the number of emitters \cite{RaimondBruneHaroche2001}.
A follow-up analysis by Andolina \etal~\cite{Andolina2019a} and Juli\`a-Farr\'e \etal~\cite{Julia-Farre2020} clarified the origin of this superextensive scaling, showing that it relies on the collective synchronisation of emitters rather than on the rapid buildup and upkeep of quantum correlations~\cite{ferraro2024reply}. This insight has important practical consequences, as it implies that the superextensive scaling can survive in systems where quantum coherence is short-lived, opening the door to room-temperature implementations in noisy environments. Indeed, the success of Ferraro's proposal, now commonly known as the \textit{Dicke quantum battery}, stems from its broad applicability, which includes---but is not limited to---cavity and circuit quantum electrodynamics~\cite{Nataf2010,Kockum2019,Crescente2020, Gemme2023}, trapped ions~\cite{DeVoe1996, Genway2014, Aedo2018, Wen2025}, semiconductor quantum dots~\cite{Scheibner2007, Raino2018, Wenniger2023}, colour centres in diamond~\cite{Bradac2017, Angerer2018, qu2025superradiance}, and fluorescent molecules~\cite{Lidzey1998, Keeling2020, Quach2022, Tibben_PRX2025, Hymas2026}.

In this chapter we focus on the latter platform, consisting of fluorescent, carbon-based molecules---also known as organic dyes---embedded in optical cavities, which underpins what are commonly referred to as \textit{excitonic quantum batteries}. The name stems from the photoactive electronic states of fluorescent molecules, known as \textit{excitons}, consisting of bound electron-hole pairs~\cite{Ostroverkhova2016}.  
Excitons, generated by the absorption of photons, act as the active medium of the battery, while the cavity acts as an optical resonator that couples distant molecules via the electromagnetic field, orchestrating their synchronisation. 

In most fluorescent dyes, light absorption is mediated by the electronic transition between the highest occupied molecular orbital (HOMO) to the lowest unoccupied molecular orbital (LUMO)~\cite{Ostroverkhova2016}. Since orbitals are typically characterised by total electronic spin $S=0$, the corresponding excitations are known as \textit{singlet} excitons~\cite{Fassioli2014, Mikhnenko2015, Ostroverkhova2016}. The shared spin character of ground (HOMO) and excited (LUMO) singlets allows this transition, which is typically associated with strong optical dipole moments of the order of several Debye~\cite{Corry2016}. However, this also makes singlets prone to rapid radiative decay back to the ground state, usually on nanosecond timescales~\cite{Ostroverkhova2016}. Therefore, when bare singlets are used as the active medium for both absorption and storage, this short lifetime sets the characteristic timescale over which energy can be retained, posing the main fundamental challenge to the practical application of excitonic quantum batteries~\cite{Campaioli2024}. This challenge is compounded by the cavity-mediated collective coupling between molecules, which acts as a double-edged sword: the same synchronisation that leads to superextensive absorption rates, known as \textit{superabsorption}, can also drive \textit{superradiance}, accelerating the radiative decay of the stored energy~\cite{Quach2022, Tibben_PRX2025, Hymas2026}.

\begin{figure*}
    \centering
    \includegraphics[width=\textwidth]{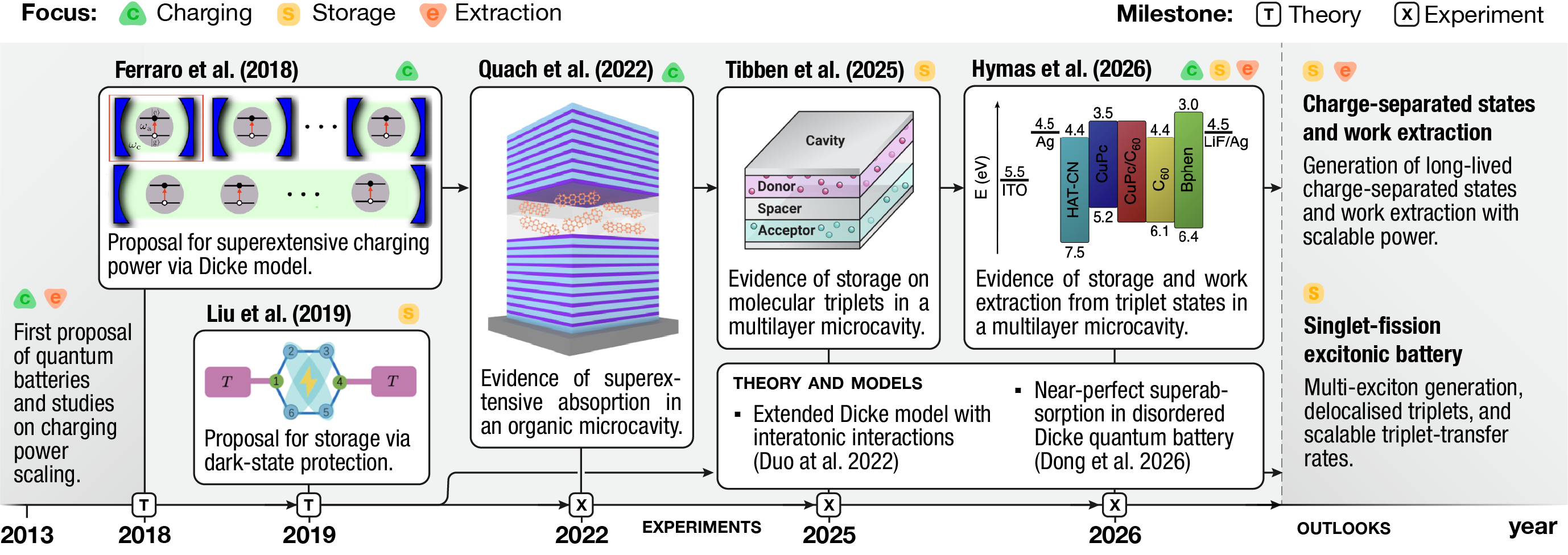}
    \caption{\textbf{Development of excitonic quantum batteries.}---Timeline of key milestones, challenges, and outlooks in excitonic quantum batteries. Quantum batteries were first proposed in the early 2010s~\cite{Alicki2013, Binder2015, Campaioli2017}, which established the link between collective effects and superextensive charging power. A pivotal milestone came in 2018 with the proposal of Ferraro \etal~\cite{Ferraro2018} for a solid-state, optically charged Dicke quantum battery, whose superextensive charging was demonstrated experimentally by Quach \etal{} in 2022 through the observation of superabsorption in an organic microcavity~\cite{Quach2022}. In parallel, early theoretical work by Liu \etal~\cite{Liu2019} addressed the storage problem via dark-state protection, exploiting the symmetry-protected states in aggregates of emitters~\cite{Quach2020}. Storage on long-lived molecular triplets was realised experimentally in 2025 by Tibben \etal~\cite{Tibben_PRX2025}, followed in 2026 by the first evidence of a full charge--storage--extraction cycle by Hymas~\etal~\cite{Hymas2026}. Throughout, theory has advanced rapidly, with new models and proposals for robust superextensive charging and storage, including extended Dicke models with inter-emitter interactions~\cite{Dou2022} and near-perfect superabsorption in disordered ensembles~\cite{Dong2026}. The principal open challenges lie in the storage and extraction stages. Near-term outlooks are the generation of long-lived charge-separated states with scalable work extraction, and the production of delocalised triplets via singlet fission with scalable transfer rates, both while preserving superextensive charging power. Figures adapted from Refs.~\cite{Ferraro2018, Liu2019, Quach2022, Tibben_PRX2025, Hymas2026}.}
    \label{fig:timeline}
\end{figure*}

A key advancement for excitonic quantum batteries came in 2022, when Quach \etal{} reported the first experimental evidence of superabsorption in an organic microcavity~\cite{Quach2022}, supporting the feasibility of superextensive charging in a solid-state platform, opening to room-temperature implementations. This milestone result brought attention to both the opportunities and challenges in excitonic quantum batteries, triggering a burst of activity from an interdisciplinary community spanning quantum information, cavity quantum electrodynamics, organic chemistry, and materials science~\cite{Campaioli2024, Camposeo2025, ferraro2026opportunities}. This concerted effort, illustrated in Fig.~\ref{fig:timeline}, has focused on developing this platform and addressing the challenge of energy storage lifetime, leading to a wealth of proposals and solutions aimed at protecting the degrees of freedom responsible for storing energy from radiative and environmental losses~\cite{Liu2019, Gherardini2020, Kamin2020,BaiAn2020, Song2024, Xu2024, Malavazi2025}.

A family of approaches exploits \textit{dark states}, i.e., excitonic states with vanishing transition dipole moment to the optical cavity mode, which are therefore decoupled from the radiation field and immune to radiative losses~\cite{Quach2020, Liu2019}. As we discuss in detail in Sec.~\ref{s:excitonic_quantum_batteries}, dark states can form via dipolar interactions in molecular aggregates, where the spatial arrangement of neighbouring molecules determines whether the individual transition dipole moments interfere constructively or destructively. This is the mechanism underlying the formation of J and H molecular aggregates, which generate delocalised excitons with, respectively, enhanced and suppressed collective transition dipole moments~\cite{Hestand2018}. In H-aggregates, destructive interference produces low-energy excitonic states that are optically dark, providing a structural route to long-lived storage, while J-aggregates concentrate oscillator strength in a bright low-energy state ideal for fast absorption---an idea that has been applied also to exciton transport~\cite{Davidson2020, Davidson2022}. This complementarity offers a natural design knob to balance fast charging against slow self-discharge, as recently explored in theoretical work on one-dimensional molecular aggregates coupled to a cavity~\cite{Li2025}.

Beyond aggregation, another approach to engineering dark states consists on exploiting molecular excitons with non-vanishing total spin states such as \textit{triplets} and \textit{quintets}, with total spin $S=1$ and $S=2$, respectively~\cite{Gallagher2015HighSpin}. Because transitions between states of different spin multiplicity are forbidden in the long-wavelength limit, triplets are typically optically dark and consequently long-lived, with lifetimes that routinely exceed milliseconds in solid-state hosts~\cite{Kabe2017, Hirata2017}, with recent groundbreaking results reporting triplet states with lifetimes exceeding 10 hours in carbon-germanium compounds at room temperature~\cite{Soto2026}. Triplets can be efficiently populated on isolated molecules via \textit{intersystem crossing}, a non-radiative spin-flip process mediated by spin-orbit coupling~\cite{Marian2021, deSilva2019InvertedGaps}, providing a natural pathway to combine fast optical charging with long lived triplet-mediated storage~\cite{Campaioli2024}. This approach, discussed in Sec.~\ref{s:molecular_triplets}, has been explored in both theoretical and experimental work, focusing on preventing triplet energy transfer from becoming a bottleneck in the overall charging process~\cite{Tibben_PRX2025, Hymas2026, Wang2025}. Here, we focus on the results of Ref.~\cite{Tibben_PRX2025}, which demonstrates a thousand-fold extension of the self-discharge time in a multilayer donor-acceptor microcavity, and Hymas \etal~\cite{Hymas2026}, who realised the first full charge-discharge cycle of an excitonic quantum battery using triplets as the metastable storage register.

Another particularly attractive route is to generate triplets via \textit{singlet fission}~\cite{Tayebjee2017, Casanova2018, Pun2019}, a spin-allowed process in which a photoexcited singlet exciton is converted into two triplet excitons residing on neighbouring molecules~\cite{Smith2010, Smith2013}, reviewed in Sec.~\ref{s:singlet_fission}. Because the process conserves total spin, it proceeds on ultrafast timescales, from tens of femtoseconds to a few picoseconds, and with minimal energy losses, outcompeting the radiative and non-radiative loss channels that limit conventional intersystem crossing~\cite{Smith2010, Musser2015, Miyata2019}. While singlet fission has been extensively studied for its applications in photovoltaics~\cite{Tayebjee2015, Trinh2015, Collins2019a} and spintronics~\cite{Smyser2020, Hudson2024framework, Casanova2018}, its use for energy storage remains largely uncharted territory. Recent work at the interface of quantum many-body theory and photochemistry has brought insights into the role of collective effects on singlet fission in extended media~\cite{MartinezMartinez2018, Climent2022, Wallner2024}, opening new avenues to promoting the rapid formation of triplet pairs while preventing their recombination into singlets~\cite{Campaioli2024_SF}. These developments suggest that the collective coupling already exploited for superextensive charging could, in principle, be repurposed to enhance triplet production rates, opening a promising path for excitonic quantum batteries.

Beyond aggregation and high-spin states, the use of \textit{charge-separated states}, discussed in Sec.~\ref{s:charge-separated} is arguably the most promising avenue to stabilise energy storage. This mechanism exploits photo-excited electron and hole pairs that are spatially separated on distinct molecular sites, rather than forming a bound exciton~\cite{Fukuzumi2014}, with a role similar to that of anion and cation of electrochemical batteries~\cite{ferraro2026opportunities}. Charge-separated states are among the longest-lived electronically excited states known in molecular systems~\cite{Ostroverkhova2016, bakulin2012role,hou2019charge}, with lifetimes that, in carefully engineered donor-acceptor architectures, range from milliseconds to seconds~\cite{Fukuzumi2014} and beyond~\cite{aprile2008longlived, tang2023manipulating}. This raises the tantalising prospect of storage times that would make excitonic quantum batteries meaningful for ordinary electronics applications. Charge separation has been studied for decades in the context of artificial photosynthesis and organic photovoltaics~\cite{Fukuzumi2014, Wasielewski2009, gust2009solar}, but its exploitation as a storage register for quantum batteries is only beginning to be considered. Notably, the full-cycle device of Hymas \etal already relies on charge-transport layers to extract energy from heterojunction~\cite{Hymas2026}, hinting at the natural compatibility of charge-separated states with this platform.

Underlying all of these proposals is the common principle of engineering \textit{metastability} in the electronic states of solid-state platforms. In each case, the goal is to channel energy from bright states that enable superabsorption, ideal for fast charging, into long-lived states that are protected from relaxation pathways. This connects excitonic quantum batteries to the broader and rapidly developing study of metastability in many-body quantum systems, currently an active research area at the interface of condensed-matter physics and quantum technology~\cite{Macieszczak2016}. Recent advances in this area have delivered methods for driving open quantum systems towards long-lived metastable states and for controlling their non-equilibrium relaxation, including anomalous relaxation phenomena such as the quantum Mpemba effect and prethermalisation~\cite{Carollo2021, Yin2024, Teza2026, Beato2026}. Beyond organic microcavities, these concepts are fundamentally platform-independent and are currently being translated for quantum energy storage applications in superconducting circuits, trapped ions, semiconductor quantum dots, and colour centres alike. 

In this chapter, we begin by reviewing molecular excitons and their coupling to optical cavity modes (Sec.~\ref{s:excitonic_quantum_batteries}). We then turn to the central challenge of energy storage, surveying proposals to transfer energy into the molecular triplet manifold (Sec.~\ref{s:molecular_triplets}), to implement singlet fission (Sec.~\ref{s:singlet_fission}), and to generate long-lived charge-separated states (Sec.~\ref{s:charge-separated}). We close with the field's most pressing challenges and promising outlooks.

\section{Excitonic quantum batteries}
\label{s:excitonic_quantum_batteries}

\subsection{Excitons in fluorescent molecules}
\label{ss:excitons-fluorescent-molecules}

\noindent
Organic fluorescent molecules, or dyes, are among the most versatile of optoelectronic materials~\cite{Ostroverkhova2016}. Fluorescent molecules absorb and re-emit light efficiently across the visible and near-ultraviolet. They are soft\footnote{Here soft is used to refer to materials held together by weak intermolecular forces, such as van der Waals interactions, $\pi$--$\pi$ stacking, and hydrogen bonds, rather than the strong covalent or ionic bonding networks of hard inorganic crystals like silicon or gallium arsenide.}, solution-processable, and chemically tunable, making them valuable in a remarkable range of settings. Dispersed at low concentration in a transparent dielectric host, they form the active layer of luminescent solar concentrators, which collect sunlight over a large area and funnel it to a small photovoltaic cell mounted at the edge of a waveguide~\cite{Debije2012,Zhang2019,Manian2021}. Blended with an electron acceptor into a bulk heterojunction, they drive charge photogeneration in organic photovoltaics~\cite{Yu1995, zhu2024progress, yang2025molecular}. Embedded between two mirrors, they couple to the confined electromagnetic field of an optical microcavity, hybridising with light to form part-light, part-matter polaritonic states~\cite{feist2018polaritonic, GarciaVidal2021, xiang2024molecular}. Arranged into molecular aggregates with controlled geometry, their collective optical response can be tuned from strongly emissive to almost dark simply through the relative orientation of neighbouring molecules, as in the red-shifted, superradiant J-aggregates of cyanine and perylene dyes~\cite{liu2010aggregate, Wurthner2011, ma2021organic}. And through chemical functionalisation, their absorption, emission, intermolecular packing, and photostability can be engineered for a given application with little change to the underlying photophysics~\cite{Anthony2006}. 

\paragraph*{Excitons.}
An exciton is a bound electron–hole pair which forms an electrically neutral excitation of an electronic system, carrying energy, spin, and momentum, but no net charge. 
In fluorescent molecules, the relative permittivity is low and the electronic wavefunctions are usually confined to a single molecule, or moiety. Due to the resulting weak dielectric screening, the electron and hole remain strongly bound, with binding energies of order $0.1$--$1\,\mathrm{eV}$, and spatially localised on the same photoactive moiety, also known as \textit{chromophore}~\cite{ghosh2020excitons}.
This is known as the \emph{Frenkel} exciton, in contrast to the weakly bound, spatially extended \emph{Wannier--Mott} exciton of conventional inorganic semiconductors, where strong screening delocalises the pair over many lattice sites~\cite{LAROCCA200397}. The Frenkel limit is what justifies treating each molecule as a self-contained quantum emitter with a discrete set of internal levels, a reduction that underpins every model in this chapter.

\paragraph*{Singlets and triplets.} The spin structure of excitons follows from the two unpaired electrons created upon excitation. The ground state is a spin singlet, with total spin $S=0$ and multiplicity $2S+1=1$, and is denoted $S_0$. Promoting one electron produces two electrons in singly occupied orbitals, whose spins may combine into either an excited singlet ($S=0$) or a triplet ($S=1$, multiplicity $2S+1=3$). Within each spin manifold the excited states are labelled by increasing energy, $S_1, S_2, \dots$ and $T_1, T_2, \dots$. The lowest triplet $T_1$ generally lies below the lowest excited singlet $S_1$, separated by the exchange splitting $E(S_1)-E(T_1) = 2K > 0$, where $E(\psi)$ denotes the energy of electronic state $\psi$ and $K>0$ is the exchange integral between the two singly occupied frontier orbitals. The splitting is a consequence of Hund's rule: the spatially antisymmetric wavefunction of the triplet keeps the two electrons apart, lowering their mutual Coulomb repulsion relative to the singlet. The ordering can, however, be inverted in carefully designed chromophores for which $2K<0$~\cite{deSilva2019InvertedGaps}.

\paragraph*{Fluorescence.}
Organic molecules couple to light through the transition dipole moment between the ground and first excited singlet~\cite{Elliot_PR1957a},
\begin{equation}
\label{eq:transition-dipole}
    \boldsymbol{\mu} = \langle S_1 | \hat{\boldsymbol{d}} | S_0 \rangle,
\end{equation}
where $\hat{\boldsymbol{d}} = -e\sum_i \hat{\boldsymbol{r}}_i$ is the electric-dipole operator, $e$ is the elementary charge, with the sum running over the electrons of the molecule at positions $\hat{\boldsymbol{r}}_i$, while $\langle\,\cdot\,|$ and $|\,\cdot\,\rangle$ denote the bra and ket of a state in Dirac notation. The brightness of the transition is quantified by the dimensionless oscillator strength $f$
\begin{equation}
    \label{eq:oscillator-strength}
    f = \frac{2 m_e \omega_0}{3\hbar e^2}\,|\boldsymbol{\mu}|^2 ,
\end{equation}
where $m_e$ is the electron mass, $\hbar$ the reduced Planck constant, and $\omega_0$ the angular frequency of the $S_0\!\leftrightarrow\!S_1$ transition, defined by $\hbar\omega_0 = E(S_1)-E(S_0)$. The same dipole moment sets both the absorption cross-section and the spontaneous (radiative) fluorescence rate $\gamma_\mathrm{r} \propto\omega_0^2 f$ for the $S_1\!\to\!S_0$ transition,
\begin{equation}
    \label{eq:radiative_fluorescence_rate}
    \gamma_{\mathrm{r}} = \frac{\omega_0^3\,|\boldsymbol{\mu}|^2}{3\pi\varepsilon_0 \hbar c^3} \;\propto\; \omega_0^3\,|\boldsymbol{\mu}|^2 ,
\end{equation}
with $\varepsilon_0$ the vacuum permittivity and $c$ the speed of light. The reciprocal $\gamma_{\mathrm{r}}^{-1}$ is the natural radiative lifetime of the $S_1$ state.
\begin{figure*}
    \centering
    \includegraphics[width=\textwidth]{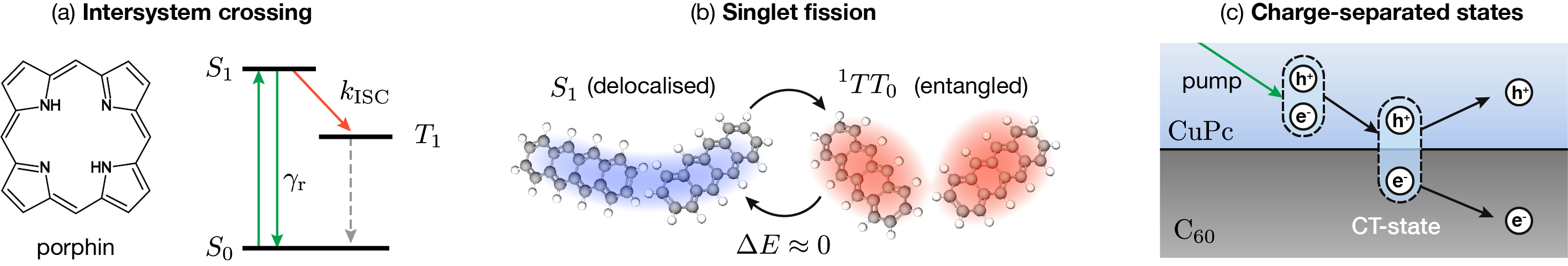}
    \caption{\textbf{Metastable states for excitonic quantum batteries.}---Three mechanisms for channelling energy from bright, optically active states into long-lived dark manifolds that protect the stored energy from radiative loss. (a) \emph{Intersystem crossing}: following absorption, the bright singlet $S_1$ either fluoresces back to the ground state $S_0$ at the radiative rate $\gamma_\mathrm{r}$ or undergoes a non-radiative transition to the lowest triplet $T_1$ at the rate $k_{\mathrm{ISC}}$, mediated by spin--orbit coupling~\cite{Marian2021, ElSayed1963}. The triplet is optically dark and consequently long-lived (dashed arrow), forming a metastable storage register~\cite{Tibben_PRX2025}. The structure shown is porphin, a prototypical chromophore exhibiting efficient intersystem crossing~\cite{Bhandari2021}. (b) \emph{Singlet fission}: a delocalised photoexcited singlet $S_1$ splits into a correlated, spin-entangled triplet pair ${}^1(TT)_0$ of singlet character, a spin-allowed and approximately energy-conserving process ($\Delta E \approx 0$) that yields two dark triplets from a single absorbed photon~\cite{Smith2010, Casanova2018, Miyata2019}. The process is illustrated for a covalently bridged tetracene dimer, an archetypal singlet-fission system~\cite{Tayebjee2017, Collins2023}. (c) \emph{Charge-separated states}: at a type-II donor--acceptor heterojunction, a pump-generated exciton in the copper phthalocyanine (CuPc) layer dissociates across the interface with C$_{60}$, first forming an interfacial charge-transfer (CT) state and then a fully charge-separated electron--hole pair~\cite{Hymas2026, Uchida_APL2004}. Such states are among the longest-lived electronic excitations in molecular systems~\cite{Fukuzumi2014, Cai2026}, offering a route to storage times relevant for practical applications.}
    \label{fig:metastable_states}
\end{figure*}

\paragraph*{Non-radiative relaxation.}
Following absorption and the formation of an $S_1$ exciton, the molecule may fluoresce back to the ground state, $S_1\!\to\!S_0$, on nanosecond timescales, or undergo \emph{intersystem crossing} (ISC) to the triplet manifold, $S_1\!\to\!T_1$. The latter is a spin-forbidden transition, enabled only by spin--orbit coupling, with a rate given, to leading order, by Fermi's golden rule,
\begin{equation}
    \label{eq:intersystem-crossing}
    k_{\mathrm{ISC}} = \frac{2\pi}{\hbar}\,\big|\langle T_1 | \hat{H}_{\mathrm{SO}} | S_1 \rangle\big|^2\,\rho_{\mathrm{FC}} ,
\end{equation}
where $\hat{H}_{\mathrm{SO}}$ is the spin--orbit coupling operator, which mixes states of different spin multiplicity, and $\rho_{\mathrm{FC}}$ is the Franck--Condon-weighted density of vibrational states of $T_1$ at the energy of $S_1$. Through this latter factor the rate acquires its characteristic dependence on the singlet--triplet gap $\Delta E_{\mathrm{ST}}$ and temperature: in the weak-coupling regime it follows the energy-gap law, decreasing approximately exponentially with increasing $\Delta E_{\mathrm{ST}}$, while the thermally activated reverse process $T_1\!\to\!S_1$ is suppressed by a Boltzmann factor $\exp(-\Delta E_{\mathrm{ST}}/k_{\mathrm B}T)$~\cite{EnglmanJortner1970}. The magnitude of the coupling matrix element is governed by El-Sayed's rules, making ISC efficient when accompanied by a change of orbital type and thus strongly enhanced by the presence of heavy atoms in the molecule~\cite{ElSayed1963, Marian2012, Marian2021}. For this reason, triplets are usually metastable and characterised by lifetimes ranging from microseconds to milliseconds~\cite{Turro2010}.

\paragraph*{Molecular aggregates.}
In a molecular aggregate---clusters of two or more molecules that are close enough to allow for exciton-exciton interactions---excitations no longer reside on isolated sites but can spatially delocalise across the entire aggregate, and singlets and triplets do so by different physical mechanisms. Optically allowed singlet excitations interact through the long-range Coulomb dipole--dipole interaction, which transfers an excitation between pairs $(i,j)$ of molecules separated by $r_{ij} = \|\boldsymbol{r}_i - \boldsymbol{r}_j\|$ with an amplitude $J_{ij}\propto|\boldsymbol{\mu}|^2/r_{ij}^3$ set by the transition dipoles and their relative orientation. Instead, dark triplet excitations carry no optical dipole moment and migrate only through the short-range exchange interaction, which decays exponentially with intermolecular separation and thus requires direct wavefunction overlap~\cite{Chen2024, Wang2024TripletManagement}. Both of these interactions induce coherent hopping when they are stronger than the coupling with the environment, leading to singlet and triplets excitation delocalisation into exciton bands whose properties depend on the sign of their couplings. Instead, when the coupling $J_{ij}$ is weak compared with the homogeneous line width, singlet transport reduces to incoherent F\"orster resonance energy transfer (FRET). This regime is appropriate to most disordered, room-temperature molecular media~\cite{Scholes2003}.

\paragraph*{J and H aggregates.}
When dipolar couplings $J$ between excitons in a molecular aggregate are larger than the homogeneous line width $\hbar\gamma$, with $\gamma$ being the total exciton relaxation rate, delocalisation changes the optical and electronic properties of the exciton bands. For simplicity, let us consider a pair of identical two-level molecules, each spanned by a ground state $|S_0\rangle$ and a singlet excited state $|S_1\rangle$ separated by $\hbar\omega_0$, with aligned transition dipoles $\boldsymbol{\mu}$. If the only interaction is the resonant dipole--dipole coupling of strength $J$, which transfers an excitation between the two sites, the collective ground and excited states, $\ket{S_0 S_0}$ and $\ket{S_1 S_1}$, respectively, are unaffected, while the two singly excited configurations $\ket{S_1 S_0}$ and $\ket{S_0 S_1}$ are mixed by
\begin{equation}
    \label{eq:dimer-hamiltonian}
    \begin{split}
    \hat{H}^{(1)} =  \hbar\omega_0 & \left( \ketbra{S_1 S_0}{S_1 S_0} + \ketbra{S_0 S_1}{S_0 S_1} \right)  \\
    + J & \left( \ketbra{S_1 S_0}{S_0 S_1} + h.c. \right),
    \end{split}
\end{equation}
where $h.c.$ denotes the Hermitian conjugate of the previous terms.
Diagonalising this single-excitation block yields the symmetric and antisymmetric eigenstates
\begin{equation}
    \label{eq:dimer-eigenstates}
    |\pm\rangle = \tfrac{1}{\sqrt{2}}\big(\ket{S_1 S_0} \pm \ket{S_0 S_1}\big), \qquad E_\pm = \hbar\omega_0 \pm J.
\end{equation}
The collective transition dipoles of the resulting states are obtained from the coherent superposition of their dipole operators 
\begin{equation}
    \boldsymbol{\mu}_\pm = \langle\pm|(\hat{\boldsymbol{d}}_1+\hat{\boldsymbol{d}}_2)|gg\rangle,
\end{equation}
and, therefore, depend on the symmetry of their wavefunction and the orientation of each individual dipole. Assuming that both dipoles are aligned and identical we obtain
\begin{equation}
    \label{eq:dimer-dipoles}
    \boldsymbol{\mu}_+ = \sqrt{2}\,\boldsymbol{\mu}, \qquad \boldsymbol{\mu}_- = 0.
\end{equation}
The symmetric state thus carries the entire, $\sqrt{2}$-enhanced, i.e., \textit{superradiant} oscillator strength and is \emph{bright}, whereas the antisymmetric state, illustrated in Fig.\ref{fig:metastable_states} together with the other metastable states considered in this chapter, is \emph{dark}, independently of the sign of $J$. What the sign of $J$ determines is which state lies lower in energy: for $J>0$ the bright state sits above the dark one, so that, after relaxation, the lowest excited state is dark and emission is suppressed---an \emph{H-aggregate}, with blue-shifted absorption---while for $J<0$ the ordering is reversed, the lowest excited state is bright, and emission is enhanced---a \emph{J-aggregate}, with red-shifted absorption~\cite{Hestand2018}. The sign of $J$ is in turn fixed by geometry through the point-dipole factor $(1-3\cos^2\theta)$, with $\theta$ the angle between the aligned dipoles and the intermolecular axis. As a result, face-to-face stacking ($\theta=90^\circ$) gives $J>0$ (H-type), and head-to-tail alignment ($\theta=0$) gives $J<0$ (J-type). This two-molecule splitting into one bright and one dark state is the seed of the extended dark manifolds that, in larger aggregates, can store excitation while remaining decoupled from the radiation field~\cite{Liu2019, Quach2020}.

\subsection{Cavity-exciton interactions}

\noindent
Let us now see how optical cavities can be used to drive collective effects in excitonic quantum batteries. For a more extensive review into optical microcavities and related structures for demonstrating strong light-matter interactions, we refer the reader to \citet{Baranov_AP2018}.
An optical cavity, or optical resonator, is a structure that traps light between reflecting boundaries~\citep{Saleh_Teich_1991}. Only electromagnetic modes whose round-trip phase satisfies the resonance condition interfere constructively and build up inside the cavity, forming a discrete set of long-lived resonant modes. By confining the field to a small volume and storing it over many optical cycles, a cavity strongly enhances the interaction between light and any matter placed inside. A single confined mode does so for all the emitters at once, irrespective of how far apart they sit, providing a common channel through which they can act collectively rather than independently.A single confined mode can do so for many emitters at once, provided they overlap appreciably with the mode profile, thus offering a common channel through which they can act collectively rather than independently. Sharing a mode in this way also changes how the system loses energy: rather than each emitter decaying through its own fast, intrinsic loss channels, the excitation is partly carried by the cavity field, and can be shielded from those individual losses when the cavity stores energy for longer than the emitters do.  
Sharing a mode in this way also modifies how the system exchanges energy with its environment: the cavity can enhance or suppress radiative decay through the Purcell effect \cite{Purcell1946}, and in the strong-coupling regime redistribute losses between light and matter according to the photonic and material fractions of the hybrid excitation~\citep{Kockum2019}.

It is this collective, light--matter character that makes the cavity a sensitive probe of interacting systems~\citep{Basov_N2021, GarciaVidal2021}, and the same principle underlies a wide range of microcavity devices, such as the semiconductor lasers of optical-fibre communications, the read and write heads of optical discs, and single-photon sources for quantum technologies~\citep{Vahala_N2003}.

\paragraph*{Fabry--P\'erot organic cavities.} Here we focus on \emph{organic cavities}, in which the active, light-emitting layer is made of organic, i.e., carbon-based,  fluorescent molecules introduced in Sec.~\ref{ss:excitons-fluorescent-molecules}. The simplest and most widely used geometry for such systems is the planar \emph{Fabry--P\'erot} cavity, in which a thin active layer is sandwiched between two parallel mirrors. These mirrors are either metallic films or dielectric distributed Bragg reflectors (DBRs), the latter being periodic stacks of transparent layers engineered to reflect a target wavelength. Their separation $L$ selects the resonant wavelengths $\lambda_m = 2nL/m$ for integer $m$, with $n$ the refractive index of the spacer. Microcavities are typically only half a wavelength thick yet centimetres across, can be fabricated by successive deposition of mirror and active layers, and operate at room temperature, making them convenient, stamp-sized platforms for strong light--matter coupling~\citep{Lidzey_2003a, Hertzog_CSR2019, Bhuyan_CR2023}. Within such microcavities, light is confined to a mode volume of order a cubic wavelength and stored for a finite time before leaking through the mirrors. This loss is measured by the quality factor $Q = \omega \tau$, corresponding to the number of optical cycles over which the field survives, and thus connected to the cavity resonant frequency $\omega$ and average photon lifetime $\tau$ in the cavity. In organic Fabry--P\'erot microcavities, $Q$ can reach several hundred to a few thousand. As we will discuss, while this is modest by the standards of optics\footnote{Quality factors of optical cavities span many orders of magnitude. Plasmonic nanogap cavities trap light in mode volumes below $1\,\mathrm{nm}^3$ but with high loss, giving $Q\sim10$~\citep{Baumberg_NM2019}; organic Fabry--P\'erot microcavities reach $Q\sim10^2$--$10^3$~\citep{Hertzog_CSR2019}; and dielectric whispering-gallery resonators such as silica microspheres and microtoroids attain $Q\sim10^8$ and beyond~\citep{Vahala_N2003}.}, it is sufficient to reach couplings that are strong enough to induce collective effects across the emitters. 

\paragraph*{Cavity field.} To describe the light-matter interaction between molecular excitons and the cavity we begin by quantizing the field inside the cavity. The latter can be decomposed it into a set of independent harmonic modes, with Hamiltonian 
\begin{equation}
    \label{eq:cavity-multimode}
    \hat{H}_{\mathrm{field}} = \sum_k \hbar\omega_k \hat{a}_k^\dagger \hat{a}_k,
\end{equation}
where $\hat{a}_k^\dagger$ creates a photon in mode $k$ of frequency $\omega_k$. When a single cavity mode of frequency $\omega_c$ is near-resonant with the molecular transition $S_0 \!\leftrightarrow\! S_1$ and well separated from the others, we can reduce the field to the single quantum harmonic oscillator,
\begin{equation}
    \label{eq:cavity-field}
    \hat{H}_{\mathrm{cav}} = \hbar\omega_c\,\hat{a}^\dagger\hat{a},
\end{equation}
whose zero-point electric field at position $\boldsymbol{r}$ has amplitude
\begin{equation}
    \label{eq:vacuum-field}
    \boldsymbol{E}_{\mathrm{vac}}(\boldsymbol{r}) = \sqrt{\frac{\hbar\omega_c}{2\varepsilon_0 V}}\,\boldsymbol{u}(\boldsymbol{r}),
\end{equation}
where $\varepsilon_0$ is the vacuum permittivity, $V$ the mode volume, and$\boldsymbol{u}(\boldsymbol{r})=\boldsymbol{\epsilon} f(\boldsymbol{r})$ is the normalized mode function, decomposed into a polarization vector $\boldsymbol{\epsilon}$ and a spatial profile $f(\boldsymbol{r})$.
\paragraph*{Strong coupling regime.} The interaction between a molecule embedded in the cavity and the confined field follows from the dipole coupling $\hat{H}_{\mathrm{int}} = -\hat{\boldsymbol{d}}\cdot\hat{\boldsymbol{E}}$, discussed in Sec.~\ref{ss:excitons-fluorescent-molecules}. Reducing a molecule to its $S_0\!\leftrightarrow\!S_1$ transition and the field to the single mode of Eq.~\eqref{eq:cavity-field}, the interaction becomes 
\begin{equation}
    \label{eq:cavity-molecule_interaction}
    \hat{H}_{\mathrm{int}} = \hbar g\,(\hat{\sigma}_+ + \hat{\sigma}_-)(\hat{a} + \hat{a}^\dagger),
\end{equation}
where $\hat{\sigma_+} := \ketbra{S_1}{S_0}$ and $\hat{\sigma_-} := \ketbra{S_0}{S_1}$, and where the single-molecule coupling reads
\begin{equation}
    \label{eq:single-coupling}
    \hbar g = -\,\boldsymbol{\mu}\cdot\boldsymbol{E}_{\mathrm{vac}}(\boldsymbol{r})
            = -\,|\boldsymbol{\mu}|\,\sqrt{\frac{\hbar\omega_c}{2\varepsilon_0 V}}\,\cos\theta\,|\boldsymbol{u}(\boldsymbol{r})|,
\end{equation}
with $\theta$ being the angle between the transition dipole $\boldsymbol{\mu}$ and the local field polarisation. The coupling is geometric and grows with the transition dipole and with field concentration (small $V$), is maximal for a dipole aligned with the field ($\theta=0$) sitting at an antinode ($|\boldsymbol{u}|=1$), and vanishes for a dipole perpendicular to the field or located at a node. When the collective exciton--cavity coupling $\sqrt{N}g$ exceeds the relevant cavity and excitonic linewidths, photons and excitons can be exchanged coherently before being lost to the environment. The corresponding cooperativity is
\begin{equation}
\label{eq:cooperativity}
C = \frac{4 N g^2}{\kappa\gamma},
\end{equation}
where $\kappa=\tau^{-1}$ is the cavity photon loss rate and $\gamma$ is the total excitonic linewidth, including radiative and non-radiative contributions. The regime $C\gg 1$ corresponds to large collective cooperativity and is commonly associated with strong light--matter coupling, in which the optical modes hybridise into polaritonic excitations~\citep{Ebbesen_ACR2016, Flick_N2018, GarciaVidal2021}.
\paragraph*{The Dicke Hamiltonian.} In organic microcavities the active layer contains not one but a large number $N$ of molecules, all sharing the same confined mode. Devices like those of Ref.~\cite{Quach2022} couple up to $N\approx 1.6\times 10^{10}$ molecular excitons to the field, with $N$ tuned over several orders of magnitude by varying the concentration of dye dispersed in the dielectric host. We are therefore interested in the \emph{collective} regime, in which the field couples to the molecules as an ensemble. Adding the dipole interaction of each molecule to the cavity field of Eq.~\eqref{eq:cavity-field} we obtain the Dicke Hamiltonian
\begin{equation}
    \label{eq:dicke}
     \begin{split} \hat{H}_{\mathrm{D}} & = \hbar\omega_c\,\hat{a}^\dagger\hat{a}
        + \sum_{i=1}^{N}\frac{\hbar\omega_0}{2}\,\hat{\sigma}_i^z
        \\ & + \hbar\sum_{i=1}^{N} g_i\big(\hat{a} 
        +\hat{a}^\dagger\big)\big(\hat{\sigma}_i^+ + \hat{\sigma}_i^-\big),
    \end{split}
\end{equation}
where $\hat{\sigma}_i^{\pm,z}$ are the Pauli operators of molecule $i$, $\omega_0$ its transition frequency, and $g_i$ its coupling to the cavity mode. The first two terms are the bare energies of the field and the molecules, and the third describes their exchange of excitations. The Dicke model is the paradigmatic description of an ensemble of two-level systems coupled to a common bosonic mode, capturing the signature of collective coupling. For identical molecules interacting via the same coupling strength $g_n = g$, the cavity couples to a \emph{single} symmetric superposition of the individual molecular states, given by the bright state
\begin{equation}
    \label{eq:bright-state}
    |B\rangle = \frac{1}{\sqrt{N}}\sum_{n=1}^{N}\hat{\sigma}_n^+|G\rangle,
    \qquad |G\rangle = |g_1 g_2 \cdots g_N\rangle,
\end{equation}
where $\ket{G}$ is the collective ground state of independent molecules. The transition between $\ket{G}$ and $\ket{B}$ couples to the cavity mode with the collectively enhanced dipolar coupling $g\sqrt{N}$ that grows with the square root of the molecular density.
This collective enhancement is the key feature for excitonic quantum batteries, accelerating the exchange of energy between light and matter and allowing the ensemble to absorb and store optical energy faster than $N$ independent molecules could.

\paragraph*{Bright polaritons.} In the single-excitation manifold, i.e., the space of states that have up to one excitation across molecules and cavity, the bright state with no photon in the cavity $|0; B\rangle:=\ket{0}_c\otimes\ket{B}$ and the single-photon state of the cavity $\ket{1; G}:=\ket{1}_c\otimes\ket{G}$, hybridise into two \emph{polaritons}, part-light and part-matter eigenstates, with energies
\begin{equation}
    \label{eq:polariton-energies}
    E_{\pm} = \hbar\omega_0 + \frac{\hbar\Delta}{2} \pm \frac{\hbar}{2}\sqrt{\Delta^2 + 4 N g^2},
\end{equation}
where $\Delta = \omega_c - \omega_0$ is the cavity-exciton detuning. 
Indeed, in this limit, the system is captured by the two-dimensional manifold spanned by $\{\ket{1;G}, \ket{0;B}\}$, associated with energies $\omega_c$ and $\omega_0$, respectively, which are coupled by the Dicke Hamiltonian via dipolar couplings with strength $g\sqrt{N}$. At resonance ($\Delta=0$) the two resulting hybrid states are called polaritons and are split by the vacuum Rabi splitting $\Omega_R = 2g\sqrt{N}$, the experimental signature of strong coupling. The upper ($+$) and lower ($-$) polaritons $\ket{P_\pm}$ each carry equal light and matter character. Only the bright state participates in this hybridisation. The photonic (light) and excitonic (matter) fraction of each polariton is given by the Hopfield coefficients, $|C_\pm|^2:=|\braket{1;G|P_\pm|1;G}|^2 $ and $|M_\pm|^2:=|\braket{0;B|P_\pm|0;B}|^2$, respectively,
\begin{align}
    \label{eq:hopfield_light}
    & |C_\pm|^2= \frac{1}{2}\left(1\pm \frac{\Delta}{\sqrt{\Delta^2 +4Ng^2}} \right) , \\
    \label{eq:hopfield_matter}
    & |M_\pm|^2= \frac{1}{2}\left(1\mp \frac{\Delta}{\sqrt{\Delta^2 +4Ng^2}} \right) ,
\end{align}
such that $|C_\pm|^2+|M_\pm|^2=1$.

\paragraph*{Dark manifold.} The remaining $N-1$ orthogonal single-excitation superpositions stay decoupled from the cavity, at the bare exciton energy $\hbar\omega_0$, and form a \emph{dark manifold} that neither absorbs nor emits into the mode. Whether a given molecular arrangement makes these collective states bright or dark is governed by the same dipole geometry that distinguishes H- and J-aggregates in Sec.~\ref{ss:excitons-fluorescent-molecules}. 

\paragraph*{Ultrastrong coupling.}  Strong coupling is reached when the collective cooperativity $C\gg 1$ (see Eq.~\eqref{eq:cooperativity}) while the normalised coupling $g\sqrt{N}/\omega_c$ measures proximity to the \textit{ultrastrong} regime in which the counter-rotating terms of Eq.~\eqref{eq:dicke}---$\hat{a}^\dagger \hat{\sigma}^+_i$ and hermitian conjugate---can no longer be neglected. In this regime, light and matter become strongly hybridized and several features beyond the standard weak- and strong-coupling descriptions may become relevant~\cite{Kockum2019}. Because organic dyes combine large transition dipoles with enormous $N$, both regimes are accessible at room temperature, with Rabi splittings of hundreds of meV~\citep{Hertzog_CSR2019, GarciaVidal2021, Bhuyan_CR2023}. The coexistence of a superradiant bright polariton, ideal for rapid optical charging, with a vast reservoir of dark states, of potential use for storage, makes organic microcavity a natural setting for an excitonic quantum battery.

\subsection{Modelling excitonic quantum batteries}
\label{ss:modelling-excitonic-qb}
\noindent
Let us begin by introducing the notation used to describe the state space of molecular excitons, which we reduced to the three electronic states relevant to charging and storage, i.e., the singlet ground state $S_0$, the first excited singlet $S_1$, and the lowest triplet $T_1$. In real molecules, each of these electronic states is dressed by a manifold of vibrational modes, and the resulting vibronic structure plays an important role in the photophysics and in the coupling to a cavity. For simplicity, here we deliberately coarse-grain this structure and model each electronic state as a single bare level, neglecting the vibrational degrees of freedom; the interested reader is referred to Ref.~\cite{Herrera2018} for a detailed treatment of the essential role of vibration--photon dressed states in nanoscale organic cavities. Every molecule is therefore a three-level system, or qutrit, with local Hilbert space $\mathcal{H}_i$ of dimension $d_i=3$, spanned by the local basis
\begin{equation}
    \label{eq:local_basis}
    \mathcal{B}_i:=\{|S_0\rangle_i,|S_1\rangle_i,|T_1\rangle_i\}.
\end{equation}

A molecule can host at most one electronic excitation, either singlet or triplet, since doubly excited configurations lie far higher in energy and are excluded from this reduced description. Excitations are thus saturable, or \emph{hard-core}. We can introduce the local singlet- and triplet-creation operators 
\begin{align}
\label{eq:singlet_creation}
    &\hat{\mathcal{S}}_i:=\mathbb{1}_1\otimes\cdots\otimes\ketbra{S_1}{S_0}_i\otimes\cdots\mathbb{1}_N, \\
    \label{eq:triplet_creation}
    &\hat{\mathcal{T}}_i:=\mathbb{1}_1\otimes\cdots\otimes\ketbra{T_1}{S_0}_i\otimes\cdots\mathbb{1}_N,
\end{align}
to express saturability via $\hat{\mathcal{S}}_i^\dagger \hat{\mathcal{T}}_i^\dagger = \hat{\mathcal{T}}_i^\dagger \hat{\mathcal{S}}_i^\dagger = \hat{\mathcal{S}}_i^\dagger \hat{\mathcal{S}}_i^\dagger = \hat{\mathcal{T}}_i^\dagger \hat{\mathcal{T}}_i^\dagger = 0$, while operators on distinct molecules commute. 
Using this basis, we can express the Hamiltonian $\hat{H}_\mathrm{ex}$ of the molecular excitons as
\begin{equation}
    \label{eq:battery-hamiltonian}
    \hat{H}_\mathrm{ex} = \sum_{i=1}^{N}\left( \hbar\omega_S\, \hat{\mathcal{S}}_i^\dagger \hat{\mathcal{S}}_i^\phdagger + \hbar\omega_T\, \hat{\mathcal{T}}_i^\dagger \hat{\mathcal{T}}_i^\phdagger \right), 
\end{equation}
with $\hbar\omega_S = E(S_1)-E(S_0)$ and $\hbar\omega_T = E(T_1)-E(S_0)$ being the singlet and triplet excitation energies introduced in Sec.~\ref{ss:excitons-fluorescent-molecules}.

\paragraph*{Battery and charger.} Following the paradigm of the Dicke quantum battery introduced by Ferraro et al. in Ref.~\citep{Ferraro2018} and developed in Refs.~\cite{Andolina2018, Andolina2019, Andolina2019a, Julia-Farre2020, Crescente2020, Dou2022, Zhang2023, Quach2022, Dong2026, Wen2025, Seidov2024, Gemme2024a, Erdman2024, Li2025, Zhang2023, Zhang2023a,  Gemme2023, Pokhrel2025, Yang2024a, Yang2024, Canzio2025, Ferraro2020, Hymas2026, Dias2026}, the organic cavity device is partitioned into a \emph{battery} ($V$), given by the ensemble of molecules in which energy is stored, and a \emph{charger} ($C$), the confined cavity mode that delivers energy from an external drive to the battery, 
\begin{equation}
    \label{eq:total_hamiltonian}
    \hat{H} = \hat{H}_C + \hat{H}_B + \hat{H}_{BC}, 
\end{equation}
with $\hat{H}_C = \hbar\omega_c \hat{a}^\dagger \hat{a}$, $\hat{H}_B = \hat{H}_\mathrm{ex}$ from Eq.~\eqref{eq:battery-hamiltonian}, and
where $H_{BC}$ represents the interaction terms between battery and charger.
Since only the optically bright $S_0\!\leftrightarrow\!S_1$ transition couples to the field, due to triplets being dark (Sec.~\ref{ss:excitons-fluorescent-molecules}), the charger and its coupling to the battery read
\begin{equation}
    \label{eq:interaction}
    \hat{H}_{BC} = \hbar\sum_{i=1}^{N} g_i \left(\hat{a}^\dagger + a\right) \left(\hat{\mathcal{S}}_i^\dagger + \hat{\mathcal{S}}_i^\phdagger\right).
\end{equation}
Charging has been studied for this system in different regimes, from the coherent dynamics of an initial $m$-photon cavity state $\ket{\psi_0} = \ket{m}_c\otimes\ket{G}$~\cite{Ferraro2018}, to incoherent pumping in resonance with the cavity mode or the polaritons~\cite{Quach2022, Han2025, Tibben_PRX2025, Dias2026, Hymas2026}.
The focus is on the rate at which energy is injected into the battery or, in case of strong couplings, the bright polaritons, which is a measure of the charging power $P = dE/dt$. Collective enhancements like bright-state couplings can accelerate this rate, so that the ensemble charges superextensively, i.e., with power $P\propto N^\alpha$ for $\alpha>1$ scaling with $N$ faster than independent molecules would~\citep{Ferraro2018, Quach2022}. Note that the triplet manifold does not participate in this energy exchange, but provides a reservoir into which the absorbed energy can be transferred and stored, as discussed below.

\paragraph*{Figures of merit.} Several figures of merit have been considered to characterise the performance of quantum batteries, focusing on power, extractable work, work fluctuations, precision, stability, energy storage lifetime, and more~\cite{Campaioli2024}. The \textit{stored energy} is typically evaluated as the change in the battery's mean energy relative to its initial state,
\begin{equation}
    \label{eq:stored-energy}
    E(t) = \mathrm{Tr}[\hat{H}_B \rho_B(t)] - \mathrm{Tr}[\hat{H}_B \rho_B(0)],
\end{equation}
with $\rho_B(t) = \tr_C\rho(t)$ being the state of the battery at time $t$.
The \textit{charging power} defined as the time derivative of the battery's mean energy,
\begin{equation}
    \label{eq:charging_power}
    P(t) = \dot{E}(t):=\frac{dE(t)}{dt},
\end{equation}
with the average power $\bar{P}(\tau) = E(\tau)/\tau$ over a charging time $\tau$ often serving as a proxy for charging speed~\citep{Ferraro2018, Binder2015, Campaioli2017}. However, not all stored energy is recoverable, either via reversible or irreversible operations, depending on the physical constraints on the controllability of the system. The maximum work extractable by cyclic unitary operations is the \emph{ergotropy}~\citep{Allahverdyan2004, Alicki2013}
\begin{equation}
    \label{eq:ergotropy}
    \mathcal{W}(t) = \mathrm{Tr}[\hat{H}_B \rho_B(t)] - \min_{U}\mathrm{Tr}\big[\hat{H}_B\, U\rho_B(t) U^\dagger\big],
\end{equation}
where the minimisation over the space of unitaries $U$ defines the passive state of $\hat{H}_B$ with respect to $\rho_B(t)$, given by the spectrum of $\rho_B(t)$ arranged with decreasing population over the increasing energy levels of $\hat{H}_B$. Because dissipation and decoherence render $\rho_B$ mixed, the ergotropy generally falls below the stored energy. Ergotropy can also be generalised and evaluated for irreversible work extraction~\cite{Lobejko2022}, i.e., by means of some operation that changes the entropy of the system and thus the spectrum of the battery's state $\rho(t)$. 

\paragraph*{Open-system dynamics.} In practice, microcavities are neither closed nor fully coherent systems. The cavity leaks photons at some rate $\kappa$, while the molecular excitons decay and dephase, and the device is continuously driven during charging. Under the Born--Markov and secular approximations, the dynamics $ \dot{\rho} = \mathcal{L}[\rho]$ of the state $\rho$ of full battery-charger system is often described with the Gorini-Kossakowski-Sudarshan-Lindblad (GKSL) master equation~\citep{Breuer2002, Milz2017, Campaioli2024a}
\begin{equation}
    \label{eq:lindblad}
    \dot{\rho} = -\frac{i}{\hbar}[\hat{H},\rho] + \sum_j \left( \hat{L}_j \rho \hat{L}_j^\dagger - \tfrac{1}{2}\{\hat{L}_j^\dagger \hat{L}_j, \rho\} \right),
\end{equation}
where each jump operator $\hat{L}_j$ encodes a dissipative channel at its characteristic rate. For instance, photon loss through the mirrors is modelled via $\hat{L}_\kappa = \sqrt{\kappa}\,\hat{a}$. For each molecule $n$, the electronic processes of Sec.~\ref{ss:excitons-fluorescent-molecules} become local jump operators. The decay of the singlet to the ground state (radiative and non-radiative) can be modelled with local jump operators $\hat{L}^{S}_i = \sqrt{\gamma_S}\,\hat{\mathcal{S}_i}$, with associated rate $\gamma_S$. Intersystem crossing of population from $S_1$ into the  triplet state $T_1$ at the rate $k_{\mathrm{ISC}}$ of Eq.~\eqref{eq:intersystem-crossing} can be modelled via local jump operators
\begin{equation}
    \label{eq:isc-jump}
    \hat{L}^{\mathrm{ISC}}_i = \sqrt{k_{\mathrm{ISC}}}\; \hat{\mathcal{T}}_i^\dagger \hat{\mathcal{S}}_i^\phdagger.
\end{equation}
Similar operators can be used to describe other local transitions, such as internal conversion $T_1 \to S_0$, representing triplet relaxation at some rate $\gamma_T$. The separation of timescales $\gamma_T \ll \gamma_S$ allows the triplet to retain energy long after the bright singlet has decayed, and it is the central mechanism exploited in Sec.~\ref{s:molecular_triplets}.

\paragraph*{Effective and exact solutions.}
Outside of some special cases that admit analytic solutions, solving Eq.~\eqref{eq:lindblad} exactly is intractable already for the number $N \sim 10^{6}$--$10^{10}$ of molecules in a real device, whose full Hilbert space grows exponentially in $N$. Often, however, the dynamics within the relevant energy manifold is captured by a small subset of degrees of freedom. This observation underlies a variety of techniques for reducing the complexity of the problem. In the low-excitation regime, characteristic of weak pumping of the cavity, the dynamics closes on the bright collective mode and the cavity, reducing the many-molecule problem to an effective single-body few-level model analogous to a Jaynes--Cummings model~\cite{Shore1993, Zhu2016, Larson2021}. More generally, when the molecules are identical and identically coupled, permutational symmetry collapses the relevant state space from exponential to polynomial in $N$, which can be exploited to integrate the master equation \textit{exactly}~\citep{Shammah2018}; numerical integrators based on this symmetry are available in software packages such as QuTiP's Permutationally Invariant Quantum Solver (PIQS)~\cite{qutip5}.

\paragraph*{Mean-field approximation.} 
For the large-$N$ regime characteristic of organic cavities, the mean-field approximation (MFA) is often used instead. It factorises the full state of the system into a product of cavity and single-molecule states, \begin{equation} \label{eq:mean-field} \rho(t) = \rho_C(t) \otimes \bigotimes_{i=1}^N \rho_{B,i}(t), \end{equation} which is equivalent to factorising all field--molecule and molecule--molecule expectation values, e.g. $\langle \hat{a}\,\hat{\sigma}^z_i \rangle \to \langle \hat{a} \rangle \langle \hat{\sigma}^z_i \rangle$, thereby discarding all connected correlations~\cite{Kusmierek2023}. The field is then replaced by its coherent amplitude $\langle \hat{a} \rangle$, and the master equation reduces to a closed set of nonlinear equations for single-operator expectation values~\citep{Kirton_AQT2019}. For a Tavis--Cummings-type model driven by incoherent pumping these read
\begin{align}
\partial_t \langle \hat{a} \rangle ={}& -\Bigl( i\omega_C + \tfrac{\kappa - \gamma_\uparrow}{2} \Bigr) \langle \hat{a} \rangle
    - i \sum_{i=1}^{N} g_i \langle \hat{\sigma}^-_i \rangle ,
    \label{eq:mf-a} \\
\partial_t \langle \hat{\sigma}^-_i \rangle ={}& -\Bigl( i\omega_M + \tfrac{\gamma_0}{2} \Bigr) \langle \hat{\sigma}^-_i \rangle
    + i g_i \langle \hat{a} \rangle \langle \hat{\sigma}^z_i \rangle ,
    \label{eq:mf-sm} \\
\partial_t \langle \hat{\sigma}^z_i \rangle ={}& \, 2 i g_i
    \bigl( \langle \hat{a}^\dagger \rangle \langle \hat{\sigma}^-_i \rangle
         - \langle \hat{a} \rangle \langle \hat{\sigma}^+_i \rangle \bigr)
    - \gamma_\downarrow \bigl( \langle \hat{\sigma}^z_i \rangle + 1 \bigr) ,
    \label{eq:mf-sz}
\end{align}
where $\gamma_0$ is the total transverse decoherence rate, $\gamma_\downarrow$ the single-molecule decay rate, and $\gamma_\uparrow$ the incoherent pumping rate of the cavity. Mean-field theory is exact in the limit $N \to \infty$ at fixed collective coupling $g\sqrt{N}$ \emph{provided} the connected correlations remain suppressed as $1/N$ throughout the dynamics. In this setting, MFA correctly reproduces superradiance and the polariton mean energies~\cite{Carollo2021b, Carollo2024}. 

\paragraph*{Breakdown of mean-field approximation.} The MFA can fail even at arbitrarily large $N$~\cite{Jin2016}. First, near dynamical or dissipative phase transitions, fluctuations grow with $N$ and the factorisation in Eq.~\eqref{eq:mean-field} breaks down in a critical region around the threshold~\cite{Minganti2018, Casteels2017}. Second, processes seeded by quantum fluctuations rather than by coherent fields are missed entirely. For an initially inverted ensemble with $\langle \hat{a} \rangle = \langle \hat{\sigma}^-_i \rangle = 0$, Eqs.~\eqref{eq:mf-a}--\eqref{eq:mf-sz} remain trapped at an unstable fixed point and the superradiant burst, triggered in reality by spontaneous emission, never develops~\cite{GrossHaroche1982}. Third, observables that are themselves correlation functions, such as incoherent photon populations $\langle \hat{a}^\dagger \hat{a} \rangle - |\langle \hat{a} \rangle|^2$, intermolecular coherences $\langle \hat{\sigma}^+_i \hat{\sigma}^-_j \rangle$, and in particular the ergotropy and entanglement measures central to this work, vanish identically under Eq.~\eqref{eq:mean-field}, so no limit in $N$ can restore them~\cite{Boneberg2022}. Systematic improvements truncate the hierarchy of equations at higher order instead, e.g., second-order cumulant expansions that retain two-operator connected correlators~\citep{Kirton_AQT2019, Plankensteiner2022}, at the price of a polynomial growth in the number of equations.

\paragraph*{Beyond mean field: cumulant expansion.} A systematic route beyond mean field is the cumulant expansion, which organises correlations by their connected (cumulant) part, e.g. $\langle \hat{A}\hat{B} \rangle_c = \langle \hat{A}\hat{B} \rangle - \langle \hat{A} \rangle \langle \hat{B} \rangle$ at second order. Truncating the hierarchy at order $n$ amounts to evolving all expectation values of up to $n$ operators exactly while setting all higher connected correlators to zero; at second order this means closing the equations with the factorisation~\cite{Kusmierek2023} \begin{equation} \label{eq:cumulant}\begin{split} \langle \hat{A}\hat{B}\hat{C} \rangle  \simeq  &\langle \hat{A}\hat{B} \rangle \langle \hat{C} \rangle + \langle \hat{A}\hat{C} \rangle \langle \hat{B} \rangle + \langle \hat{B}\hat{C} \rangle \langle \hat{A} \rangle \\
&- 2 \langle \hat{A} \rangle \langle \hat{B} \rangle \langle \hat{C} \rangle\end{split}\end{equation} Mean field, Eqs.~\eqref{eq:mf-a}--\eqref{eq:mf-sz}, is recovered as the first order of this hierarchy. At second order the dynamics of, e.g., the photon number $\langle \hat{a}^\dagger \hat{a} \rangle$ couples to the field--molecule coherences $\langle \hat{a}^\dagger \hat{\sigma}^-_i \rangle$, whose equations of motion are in turn closed via Eq.~\eqref{eq:cumulant}~\cite{Kirton2018}. This retains precisely the two-operator connected correlations discarded by Eq.~\eqref{eq:mean-field}---incoherent photon populations, fluctuation-seeded superradiance, and the leading contributions to collective coherence---at the cost of a number of equations that grows polynomially with the number of distinct molecular classes, and only as $O(1)$ for identical molecules, where permutational symmetry reduces all correlators to a few representative ones~\citep{Kirton_PRL2017}. Symbolic derivation and integration of cumulant equations to arbitrary order is automated in the Julia package QuantumCumulants~\citep{Plankensteiner2022}. The truncation is uncontrolled in the sense that no small parameter guarantees convergence order by order; it is accurate when higher cumulants are suppressed (typically as higher powers of $1/N$), but can fail in strongly correlated or critical regimes, where the hierarchy may even develop unphysical solutions.

\paragraph*{Beyond mean field: Tensor networks.} 
When correlations must be tracked without truncating at a fixed order, tensor-network methods provide a variational compression of the many-body state~\cite{Montangero2018}. A matrix-product state (MPS) ansatz \begin{equation} \label{eq:mps} |\psi\rangle = \sum_{s_1, \ldots, s_N} A^{[1] s_1} A^{[2] s_2} \cdots A^{[N] s_N} \, |s_1 \ldots s_N\rangle , \end{equation} with matrices $A^{[i]s_i}$ of bond dimension $\chi$, parametrises states whose bipartite entanglement entropy is bounded by $\log \chi$, reducing the cost of time evolution to polynomial in $N$ and $\chi$. Mixed states are treated analogously by vectorising $\rho$ into a matrix-product density operator evolved under the Lindbladian. The all-to-all connectivity of the light--matter problem, however, lacks the one-dimensional structure these methods were designed for, and only a few ansätze remain controlled: arranging the molecules into an effective chain with the cavity at its boundary (justified because all intermolecular correlations are mediated by the cavity), star-to-chain transformations of the bath, and tree tensor networks whose hierarchical geometry accommodates the collective coupling. Mature implementations of these algorithms are available in libraries such as ITensor~\citep{Fishman_SciPost2022} and Quantum TEA~\citep{Ballarin_qtealeaves}. A complementary use of the same compression applies along the time direction: for molecules strongly coupled to structured vibrational environments, the non-Markovian influence functional can be represented as a matrix-product operator over time steps, as in the time-evolving matrix product operator (TEMPO) algorithm~\citep{Strathearn_NC2018} and its process-tensor generalisations~\citep{Jorgensen_PRL2019}. These methods are implemented in OQuPy~\citep{Fux_JCP2024}, which moreover combines process tensors with the mean-field reduction of Eq.~\eqref{eq:mean-field} to treat large ensembles of vibrationally dressed molecules coupled to a common cavity mode, retaining non-Markovian single-molecule physics exactly while factorising only the molecule--photon correlations.

\paragraph*{Exact solvability via integrability.} 
A separate route to handling Eq.~\eqref{eq:dicke} exactly, distinct from the numerical methods above, exploits the integrable structure of the model itself. Under the rotating-wave approximation, the Dicke Hamiltonian reduces to the Tavis--Cummings (TC) Hamiltonian, in which the counter-rotating terms $\hat a^\dagger\hat\sigma_i^+$ and $\hat a\hat\sigma_i^-$ are dropped. This is not merely a simplification: it restores a continuous $U(1)$ symmetry, generated by the total excitation number $\hat M = \hat S^z + \hat a^\dagger \hat a$, which commutes with the Hamiltonian and is broken by the counter-rotating terms present in the full Dicke model. \citet{Bogolyubov_JOMS2000a} showed that this conserved quantity, together with the total-spin Casimir operator, allows the many-body TC Hamiltonian to be diagonalised exactly for arbitrary $N$ using the algebraic Bethe ansatz, the same quantum-inverse-scattering machinery underlying the Gaudin model. Eigenstates take the form of Bethe vectors built from a set of auxiliary spectral parameters, or rapidities, whose values are fixed by a system of coupled nonlinear equations, the Bethe equations; the corresponding eigenenergies follow as simple linear combinations of these rapidities. This furnishes a genuinely exact solution of the many-body problem within each excitation-number sector, in contrast to the permutationally-symmetric numerics discussed above, which remain a numerical method even though they exploit symmetry to make the calculation tractable.

It is natural to ask whether the full Dicke Hamiltonian, prior to the rotating-wave approximation, admits an analogous treatment. The answer is more subtle than a simple yes or no. For a single emitter ($N=1$), the Dicke Hamiltonian reduces to the quantum Rabi model, which lacks the $U(1)$ symmetry of its TC counterpart entirely: the counter-rotating term breaks it down to a residual, discrete $\mathbb{Z}_2$ parity. \citet{Braak_PRL2011a} showed that this much weaker symmetry is nonetheless sufficient for exact solvability, though by an entirely different route than the Bethe ansatz: the regular spectrum in each parity sector is given by the zeros of a transcendental function constructed from a recursively-defined power series in the coupling, rather than by the roots of an algebraic equation. Braak further proposed a general criterion for quantum integrability, requiring that every eigenstate admit a unique label built from one quantum number per discrete and continuous degree of freedom, and noted explicitly that this criterion extends to the Dicke and Tavis--Cummings models. Whether a closed-form solution of comparable generality exists for the many-body Dicke Hamiltonian with $N>1$ and unbroken counter-rotating terms remains, to our knowledge, an open problem; the present absence of a known solution should not be mistaken for a proof of unsolvability.

These two results together clarify what is, and is not, lost in applying the rotating-wave approximation to Eq.~\eqref{eq:dicke}. The approximation is often justified on the grounds that the counter-rotating terms are small in the experimentally relevant near-resonant, weak-coupling regime; the analysis above shows that it does something more structural, exchanging the discrete parity symmetry of the full light--matter problem for a continuous conservation law that admits a qualitatively different and more tractable exact solution. Braak's own counterexample, an exactly solvable generalisation of the Rabi model possessing no symmetry at all, demonstrates that exact solvability and integrability are not equivalent notions in quantum mechanics, in contrast to the classical, Liouville-integrability setting in which the two are more closely tied. For the dissipative, driven dynamics of interest in most experimental devices, however, neither the Bethe-ansatz solution of the TC model nor the Rabi-model solution applies directly, since both are coherent, closed-system results, and one must turn to approximate methods for the open-system dynamics that follow.

\paragraph*{Beyond the Dicke model.}
The Dicke model provides a useful minimal description of excitonic quantum batteries because it isolates the bright collective excitation that couples to the cavity field. Its standard form, however, implicitly assumes a spatially uniform single-mode cavity: all molecules are taken to have the same transition energy, the same dipole orientation, and the same light--matter coupling. This is a strong idealisation. In a realistic cavity, the electromagnetic field has a spatial profile with nodes and antinodes, a local polarization, and possibly several relevant modes. The coupling of molecule $i$ to mode $\mu$ is therefore controlled by the local vacuum field,
\begin{equation}
    g_{i\mu}
    =
    -\frac{1}{\hbar}
    \boldsymbol{d}_i\cdot
    \boldsymbol{E}_{\mu,\mathrm{vac}}(\boldsymbol{r}_i),
    \label{eq:gim}
\end{equation}
so that molecules located at different positions, or with different dipole orientations, do not contribute equally to the collective bright state.

These corrections become especially important in open, lossy, or nanophotonic cavities, where the electromagnetic environment is generally multimode and dispersive. In such cases, the notion of a single mode volume and a single coupling constant is only an approximation. A more general treatment is provided by macroscopic quantum electrodynamics, where the cavity is described through the dyadic Green function $\boldsymbol{G}(\boldsymbol{r},\boldsymbol{r}',\omega)$ of Maxwell's equations~\citep{KnollScheelWelsch2001,ScheelBuhmann2008,Buhmann2012}. This formulation naturally includes spatially non-uniform fields, mode dispersion, absorption, leakage, and cavity-mediated interactions between distinct emitters. The Dicke model is recovered only when the Green function is dominated by a single spectrally isolated resonance whose field profile is approximately uniform over the molecular ensemble. Away from this limit, the effective number of molecules participating in the collective excitation may be significanlty smaller than the total number of molecules, and bright and dark excitonic sectors can be mixed by spatial disorder, multimode coupling, and dissipation~\citep{NovotnyHecht2012,TormaBarnes2015}.

\section{Storing energy in molecular triplets}
\label{s:molecular_triplets}
A major advance in Dicke quantum batteries was recently reported with the experimental observation of superabsorption in an organic optical microcavity~\cite{Quach2022}. 
The findings indicate that the rate at which energy is transferred from the cavity mode to the photoactive molecules—analogous to the charging speed of a battery—grows superextensively with the number of molecules. 
This provides strong support for the superabsorption concept, namely that devices can charge more rapidly as their storage capacity increases.

However, superabsorption comes with an inherent trade-off. 
While it enables extremely rapid charging through efficient energy transfer to the emitters, it is accompanied by its counterpart, \textit{superradiance}. 
This effect introduces a cavity-enhanced decay channel, causing accelerated energy loss and thus rapid and spontaneous discharge~\cite{Chen_PRL2013a,Crisp_NL2013a,Svidzinsky_PRX2013a,Aberra-Guebrou_PRL2012a,Garraway_PTOTRSAMPAES2011a,Prasad_PRA2010a,Temnov_PRL2005a,Prasad_PRA2000a,GrossHaroche1982,Dicke_PR1954a}. 
Although local decoherence can mitigate superradiance to some extent~\cite{Quach2022}, the intrinsic radiative lifetime of individual emitters—typically on the nanosecond scale—places a fundamental upper bound on the energy storage time in this type of Dicke quantum battery. 
Consistent with this, the initial experimental realization reported by \citet{Quach2022} exhibited very short self-discharge times on the order of nanoseconds. 
Consequently, a central challenge in this field is to identify mechanisms that can extend the storage lifetime, which is essential for practical energy storage applications.
In this section we review one such approach, based on the transfer of population (energy) to long--lived molecular triplet states.

\subsection{Generating triplets via intersystem crossing}

There are two identified ways to transfer energy from singlets to triplets in a molecular system:
(i) \textit{via} intersystem crossing, or
(ii) \textit{via} direct polariton-triplet coupling~\cite{Tibben_PRX2025}.
Here, we start by explaining the role of intersystem crossing, before moving on to polariton-triplet coupling in the next section.

Following superabsorption, energy can be directly transferred to the triplet state \textit{via} intersystem crossing.
Intersystem crossing occurs when excited state energy relaxes non-radiatively to an electronic state of different spin multiplicity, typically from a singlet state to a triplet state.
This ``spin flip'' is driven by local enhancements to the magnetic field environment, and in molecular systems is often facilitated by the heavy atom effect and its perturbation of molecular spin-orbit coupling. 
Due to Pauli exclusion, these triplet excitons are spin-forbidden from decaying directly to the singlet ground state, and form the basis of metastable storage in molecular quantum batteries.

The population of triplet states \textit{via} intersystem crossing was first systematically characterised in terms of a classical rate equation model by \citet{Mukherjee_JACS2023}, where an increased rate of intersystem crossing was observed when the lower polariton of a hybridised iodine-modified fluorescein dye and cavity mode was in resonance with the triplet state energy of the dye.
It was identified that polariton-mediated intersystem crossing can be 5 orders of magnitude faster, when the polariton and triplet energies are in resonance, than in the bare molecule outside a cavity.
As the lower polariton is detuned to lower energies than the triplet state, reverse intersystem crossing starts to dominate and funnel energy back into the hybridised states~\cite{Stranius_NC2018,Eizner_SA2019,Yu_NC2021}.

This mechanism of triplet population requires the rate of intersystem crossing $\gamma_{\mathrm{ISC}}$ to be much faster than all other competing loss channels,
\begin{equation}
    \gamma_\mathrm{ISC} > \gamma_C,\, \gamma_D,\, \gamma_A.
    \label{eq:isc_condition}
\end{equation}
In a typical molecular quantum battery, these loss channels are defined by the absorptive singlet state lifetimes ($\gamma_D^{-1}$ and $\gamma_A^{-1}$) and the cavity photon lifetime ($\gamma_C^{-1}$).

In the results of~\citet{Tibben_PRX2025}, the intersystem crossing rate was on the order of these other loss channels, resulting in triplet population \textit{via} intersystem crossing being a non-dominant pathway.
However, in the results of~\citet{Hymas2026}, the intersystem crossing rate exceeded the rate of these other loss channels, resulting in intersystem crossing being identified as the dominant triplet population pathway. 

The design of~\citet{Hymas2026} draws inspiration from a cavity photodiode design, where an absorber with high rate of intersystem crossing is sandwiched between charge transport layers to achieve effective charge separation as a photocurrent~\cite{Eizner_AP2018}.
The chosen photoabsorber, copper(II) phthalocyanine, exhibits very fast intersystem crossing ($\gamma_\mathrm{ISC}^{-1}\approx200$~fs), which outcompetes other dissipative losses in the system~\cite{Dutton_PRB2010}.
The triplet lifetime of this material is up to 50 ns, which is the storage lifetime of this device and somewhat shorter than what can be expected for most other triplet materials~\cite{Caplins_PCCP2016,McVie_JCSFT2MCP1978,Gouterman_JoMS1961,Mukherjee_CC2015}.
The reduction in triplet lifetime of this material is due to the paramagnetism of the Cu$^{2+}$ ion, which leads to the triplet state taking on a doublet character (forming a so-called "trip-doublet" state) and activating nascent non-radiative relaxation pathways, bypassing Pauli exclusion~\cite{Ma_SSC2001,Asano-Someda_IC1999,Cory_CR1991,Bruder_OE2010}. 
While negatively affecting metastability of the storage mechanism in the quantum battery, this effect also directly results in the extremely fast intersystem crossing rate in this material.
In the quantum battery device, the metastability of the copper(II) phthalocyanine was enough to facilitate efficient charge separation, allowing for ultrafast characterisation of charging and discharging dynamics~\cite{Hymas2026,Dutton_JPCC2012}. 

Broadband transient reflectance measurements (500-1000~nm, probe delays up to 16~\textmu s) confirm metastabilisation of the stored energy. Following ultrafast charging \textit{via} superabsorption, excited state energy was rapidly funnelled to the triplet manifold and persisted well into the nanosecond regime.

The separation of charge demonstrated by~\citet{Hymas2026} will be discussed in more detail in Sec.~\ref{s:charge-separated}.

\subsection{Polariton-triplet energy transfer}

In this section, we explain how triplets can be populated \textit{via} a direct coupling with the cavity. At moderate to high dye concentrations, molecular aggregation can endow triplet states with a partial singlet character \textit{via} random disorder~\cite{Sternlicht_TJoCP1963,Atkins1975,Forecast_JCTC2023}. These ``bright triplets'' acquire a non-negligible optical dipole moment, enabling direct coupling with the cavity at strengths much lower than the singlet state. When a polariton state is isoenergetic with the triplet state, a hybridisation occurs which produces a polariton-triplet admixture and opens up an avenue to direct excited state population transfer to the triplet. However, due to this hybridisation, the lifetime of the resulting hybrid state  is shortened relative to the triplet. This mechanism is therefore also a double-edged sword: the resonance that drives efficient triplet population simultaneously reduces triplet lifetime by mixing in short-lived polariton character.
\begin{figure*}
    \centering
    \includegraphics[width=0.8\linewidth]{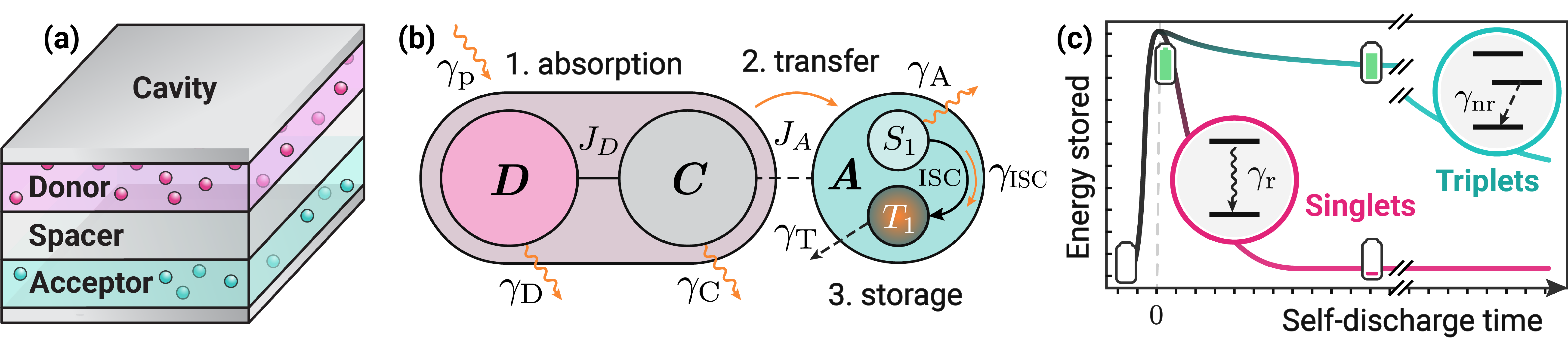}
    \caption[Microcavity--based quantum battery design and its energy dynamics]{\textbf{Microcavity--based quantum battery design and its energy dynamics.---}
    (a)~The device is based on a multilayered organic microcavity where donor (charging) and acceptor (storage) layers are spatially separated.
    (b)~Resonant pumping ($\gamma_p$) results in strong coupling between donor and cavity $J_D$, facilitating rapid charging \textit{via} superabsorption~\cite{Quach2022} to polariton states. Cavity interactions with the acceptor $J_A$ allow excited state population to relax to the metastable triplet state $T_1$ \textit{via} intersystem crossing ($\gamma_{\mathrm{ISC}}$) or polariton interactions, out--competing dissipative processes ($\gamma_\mathrm{D}, \gamma_\mathrm{C} \text{ and } \gamma_\mathrm{A}$).
    (c)~Population of the triplet state in the acceptor results in energy retention ($\gamma_{\mathrm{nr}}$) several orders of magnitude longer than in the donor singlet state ($\gamma_\mathrm{r}$).
    Adapted with permission from~\citet{Tibben_PRX2025}.}
    \label{fig:device_design}
\end{figure*}

In 2025, we proposed a proof-of-concept Dicke quantum battery device implemented in a multilayer optical microcavity~\cite{Tibben_PRX2025}, drawing on the architecture introduced by Zhong \textit{et al.}~\cite{Zhong_ACIE2017} and schematically illustrated in Fig.~\ref{fig:device_design}. 

In this design, strong light-matter coupling between a selected cavity electromagnetic mode and a designated “donor” layer gives rise to a collectively enhanced absorption process, thereby enabling superabsorption and consequently rapid battery charging, in line with the experimental demonstration by \citet{Quach2022} and as shown in Fig.~\ref{fig:device_design}b. 
The harvested excitation energy is then funnelled into an acceptor \textit{storage layer} composed of molecular species that exhibit efficient intersystem crossing and long-lived triplet states, which effectively suppresses superradiant emission channels and thus mitigates unwanted superradiant battery discharge.
This is portrayed in Fig.~\ref{fig:device_design}c, where, in the absence of an optically-accessible triplet state, the metastability of the battery would decay. 

Five devices with systematically varied cavity-mode energies were fabricated \textit{via} sequential thin-film deposition, sweeping the LP--triplet detuning $\Delta E$ from $+0.216\,\mathrm{eV}$ (Cavity~1) to $-0.091\,\mathrm{eV}$ (Cavity~5), with the Cavity 4 being closest to an isoenergetic detuning, \textit{i.e.} $\Delta E \approx 0$.
The full device detunings are listed in Table~\ref{t:parameters}.
The total mirror separation $L$ was adjusted for each device to position the second-order Fabry--P\'erot antinode at both organic layers.

\begin{table}
    \centering
        \begin{tabular*}{\columnwidth}{@{\extracolsep{\fill}}lcccc}
            \hline
            Device   & $\omega_C$ (eV) & $J_D$ (eV) & $J_A$ (eV) & $\Delta E$ (eV) \\
            \hline \hline
            Cavity 1 & 2.12 & 0.23 & 0.07 & $+$0.216 \\
            Cavity 2 & 1.97 & 0.23 & 0.10 & $+$0.094 \\
            Cavity 3 & 1.89 & 0.23 & 0.07 & $+$0.036 \\
            Cavity 4 & 1.88 & 0.25 & 0.08 & $+$0.011 \\
            Cavity 5 & 1.79 & 0.27 & 0.13 & $-$0.091 \\
            \hline
        \end{tabular*}
        \caption{Jaynes--Cummings model parameters for each device.
        $\omega_C$ is the cavity-mode energy at normal incidence, $J_D$ and $J_A$ are the cumulative cavity-donor and cavity-acceptor   couplings, and $\Delta E = E_\mathrm{LP} - E_T$ is the LP-triplet detuning.}
        \label{t:parameters}
\end{table}

Steady-state and time-resolved emission measurements across the five devices reveal that energy transfer to the storage layer proceeds \textit{via}  optically-driven polariton-triplet resonance, rather than \textit{via} intersystem crossing. 
The intrinsically inefficient ISC of R6G rules out this energy transfer \textit{via} ISC in this device, and a fit to the relative change in fluorescence intensity across the detuning sweep yields a weak triplet-cavity coupling $J_T\approx5$meV, consistent with the small residual dipole moment expected of aggregation-induced bright triplets. 
At near-zero detuning, Cavity~4 reaches a triplet population of approximately 50\% and a self-discharge time of 40.3 $\pm$ 0.4\,$\mu$s, a thousand-fold extension over the nanosecond-scale self-discharge previously reported by~\citet{Quach2022}.

This result, however, carries an intrinsic limitation that mirrors the chapter's central tension at a smaller scale. The very polariton-triplet hybridisation that allows the triplet manifold to be populated efficiently also partially borrows its lifetime from the comparatively short-lived polariton, since $$\tau_{\tilde T}\approx p_T\tau_T+p_P\tau_P;$$ the closer the device is tuned to resonance, the more efficiently the triplet is populated, but the more its metastability is eroded by admixture with the bright state responsible for charging it in the first place. Indeed, the reported storage time is shorter than the bare PdTPP triplet lifetime measured in the absence of a cavity, and only a narrow window of detuning sustains both efficient transfer and a non-negligible storage time. Mechanism~1, in which intersystem crossing populates a triplet state fully decoupled from the polariton, evades this lifetime trade-off in principle, and it is precisely this regime that is realised in the device of \citet{Hymas2026},  which we discuss later in Sec.~\ref{s:charge-separated}. 

\section{Generating triplets via singlet fission}
\label{s:singlet_fission}

\noindent 
Let us now look at singlet fission, another approach to generate metastable excitations in organic semiconductors. Singlet fission is a spin-allowed photophysical process in which a photoexcited singlet exciton, localised on one chromophore or delocalised across multiple ones, splits into a pair of triplet excitons, so that a single absorbed photon yields two electronic excitations~\cite{Smith2010,Smith2013}. In a pair of coupled molecules, singlet fission is often schematically written as $S_0 S_1 \!\rightarrow\! {}^1(T_1T_1) \!\rightarrow\! T_1 + T_1$, corresponding to the splitting of a bright singlet $S_0 S_1$ into a correlated triplet pair ${}^1(T_1T_1)$ with \textit{singlet character}, and its subsequent separation into two independent triplets $T_1 + T_1$~\cite{Miyata2019,Casanova2018}. Because each triplet $T_1$ carries roughly half the energy of the singlet $S_1$, SF is a molecular form of carrier multiplication that is, in principle, near-unity efficient and operative on ultrafast timescales~\cite{Smith2010,Chan2013}.

\paragraph*{Singlet fission for energy conversion.} The original and still dominant motivation for studying SF is photovoltaics, where exciton multiplication offers a route to circumvent the single-junction Shockley--Queisser limit~\cite{shockley1961}. By converting one high-energy photon into two charge carriers, SF absorbers can in principle raise the detailed-balance efficiency from $\sim$33\% to $\sim$44\%~\cite{Hanna2006}. This promise has been demonstrated experimentally. Fission-based organic photovoltaic cells using pentacene reach external quantum efficiencies\footnote{In photovoltaics (PV), External Quantum Efficiency (EQE) is defined as the ratio of the number of charge carriers (electrons) collected by the solar cell’s external circuit to the number of incident photons striking the cell at a specific wavelength. Multiple exciton generation processes allow EQE to exceed 100\%~\cite{semonin2011meg}.} above 100\%~\cite{congreve2013}, and tetracene interlayers have been used to sensitise crystalline silicon, with a combined fission-plus-transfer yield approaching 133\%~\cite{Einzinger2019,Macqueen2018}. These results establish SF as a mature strategy for boosting the photon-to-excitation yield of an absorber~\cite{rao2017,casillas2020}.

\paragraph*{Spectral conversion.} Taken together with its reverse process---known as triplet fusion or triplet-triplet annihilation---SF underpins a family of photon-management schemes that are directly relevant to energy storage and transduction. The production of two excitons per absorbed photon (down-conversion of a high-energy photon into two lower-energy excitations), photon up-conversion through triplet fusion~\cite{Gholizadeh2020, Forecast_JCTC2023}, and, more generally, spectral transduction between the energy at which light is absorbed and the energy at which excitations are stored or re-emitted~\cite{rao2017,Smith2013,casillas2020}. 

\paragraph*{Spintronics and molecular logic.} The value of SF, however, extends well past charge generation, because the correlated triplet pair is a doubly excited, maximally spin-entangled state with a rich internal spin structure~\cite{Smyser2020,Scholes2015}. Time-resolved electron spin resonance has revealed long-lived quintet pair states and exchange-coupled triplet pairs that survive for hundreds of nanoseconds~\cite{Tayebjee2017,Weiss2017,Bayliss2018}. Such optically prepared, spin-polarised states are attractive for spintronics and quantum information, since they can be addressed coherently, manipulated with magnetic-resonance pulses to implement room-temperature quantum gates, and used as a source of entangled spin pairs and spin-based logic~\cite{Smyser2020,Dill2023,Bardeen2019,Hudson2024framework}. In this sense SF is not only a way to multiply excitons, but also a chemically tunable factory for dark, high-spin quantum states.

\paragraph*{The challenges.} Although singlet fission and triplet fusion are well understood in weakly interacting systems or small strongly-interacting aggregates like molecular dimers, characterising and optimising singlet fission in an extended and strongly coupled material is still a challenge~\cite{Miyata2019}. This is because the process is generally not a purely electronic transition, but rather the result of interplay of excitonic, vibrational and optical degrees of freedom that are coupled on overlapping, femtosecond-to-nanosecond timescales~\cite{Miyata2019}. Vibronic coupling and conical intersections drive the primary step, so the nuclear motion cannot be reduced to simple memory-less, i.e., Markovian, reservoir. The system is dressed by a structured phonon bath, and the resulting dynamics are markedly non-Markovian, with bath memory and coherent vibronic feedback shaping both the population and the spin evolution~\cite{Schnedermann2019, Alvertis2019}. Compounding this, the correlated multiexciton state is intrinsically entangled---in its spin structure and between the excitons and the vibrations that mediate fission---so reduced single-particle or rate-based descriptions are frequently inadequate, and a faithful treatment must retain system--bath correlations explicitly~\cite{Miyata2019,Smyser2020,Scholes2015}. Set against the disorder and morphological heterogeneity of real films, disentangling the singlet, triplet and quintet contributions to this coupled exciton--phonon--photon dynamics remains a central open problem~\cite{Miyata2019,Casanova2018}. A second, practical bottleneck is the extraction step. Harvesting the triplets requires efficient, typically short-range Dexter transfer across an interface, whether to silicon or to a nanocrystal acceptor, and the transfer rate is sensitive to interlayer chemistry and passivation~\cite{Einzinger2019,Tabachnyk2014,Allardice2019,Davis2018, Baldacchino2022,Baldo2025scalablepathway,Baldacchino2025beyondtetracene}. Optimising triplet capture at the interface with silicon, in particular, has required atomically thin engineered interlayers~\cite{Einzinger2019,Macqueen2018}.

\subsection{Spin structure of the triplet pair}

\noindent 
Before introducing a model for exploiting SF in excitonic quantum batteries in Sec.~\ref{ss:sf-battery-model}, we review the basic model for singlet fission in molecular dimers.
Let us consider a two-molecule system in which each chromophore, $A$ and $B$, supports a ground state $S_0$, a bright singlet exciton $S_1$, and a triplet. Each triplet is a spin-1 system, spanned by the three sub-levels $\ket{1,m}_i$ with $m \in \{-1,0,+1\}$, the eigenstates of the local spin projection $\hat S_{z,i}$~\cite{Collins2023}. The spin structure of a single triplet $i$ is governed by two interactions: a zero-field splitting (ZFS), arising from the magnetic dipolar coupling of the two unpaired electrons, and a Zeeman coupling to an applied magnetic field $\mathbf B$. In the principal-axis frame of triplet $i$ these read 
\begin{align}
    \label{eq:zfs}
    &\hat{h}_\mathrm{zfs}^{(i)} = D_i\!\left(\hat S_{z,i}^{2}-\tfrac{2}{3}\right) + E_i\!\left(\hat S_{x,i}^{2}-\hat S_{y,i}^{2}\right), \\
    \label{eq:zeeman}
    &\hat{h}_\mathrm{z}^{(i)} = 
    g\mu_B\,\mathbf B\!\cdot\!\hat{\mathbf S}_i,
\end{align}
with $D_i,E_i$ the ZFS parameters set by the molecular geometry (and generally non-collinear between $A$ and $B$), and with $g$ being the $g$-factor and $\mu_B$ the Bohr magneton~\cite{Collins2023,Bayliss2016}. When the two triplets are brought together they interact through the Heisenberg exchange coupling,
\begin{equation}
    \label{eq:exachange}
    \hat{H}_{\mathrm{ee}} = J\,\hat{\mathbf S}_A\!\cdot\!\hat{\mathbf S}_B,
\end{equation}
here chosen to be isotropic for simplicity, whose strength $J$ reflects the inter-triplet wavefunction overlap~\cite{Merrifield1971,Collins2023}. Interacting triplet pairs are described by nine possible basis states which, in the total-spin basis, for three manifolds: a single overall-\emph{singlet} pair state, three \emph{triplet} states, and five \emph{quintet} states. These are the simultaneous eigenstates of the total-spin operators $\hat{\mathbf S}^2$ and $\hat S_z$, with $\hat{\mathbf S}=\hat{\mathbf S}_A+\hat{\mathbf S}_B$~\cite{Merrifield1971,Smyser2020,Collins2023}. We denote these manifolds by $\ket{{}^1(TT)}$, $\ket{{}^3(TT)_m}$ and $\ket{{}^5(TT)_m}$. Their existence is not merely formal: long-lived quintet pair states produced by fission have been observed directly by transient electron-spin-resonance spectroscopy~\cite{Pun2019,Tayebjee2017,Weiss2017,Bayliss2018}.

\paragraph*{The triplet-pair spin Hamiltonian.} Collecting these terms, the spin dynamics of the correlated triplet pair is captured by the Hamiltonian~\cite{Collins2023,Smyser2020}
\begin{equation}
\label{eq:Htt}
\hat H_{\mathrm{TT}} = \hat H_{\mathrm{z}} + \hat H_{\mathrm{zfs}} + \hat H_{\mathrm{ee}},
\end{equation}
the sum of Zeeman, zero-field-splitting and exchange contributions,
\begin{equation}
\label{eq:Httterms}
\begin{split}
\hat H_{\mathrm{z}}   ={}& g\mu_B\,\mathbf B\!\cdot\!\big(\hat{\mathbf S}_A+\hat{\mathbf S}_B\big),\qquad
\hat H_{\mathrm{ee}} = J\,\hat{\mathbf S}_A\!\cdot\!\hat{\mathbf S}_B, \\
\hat H_{\mathrm{zfs}} ={}& \sum_{i=A,B}\!\Big[\, D_i\!\left(\hat S_{z,i}^{2}-\tfrac{2}{3}\right)
   + E_i\!\left(\hat S_{x,i}^{2}-\hat S_{y,i}^{2}\right)\Big].
\end{split}
\end{equation}
The exchange term is diagonal in total spin, with $\hat{\mathbf S}_A\!\cdot\!\hat{\mathbf S}_B=\tfrac12\big[S(S+1)-4\big]$, so that the singlet, triplet and quintet manifolds are split by energies $\{-2J,-J,+J\}$~\cite{Smyser2020,Bayliss2016}. Crucially, both $\hat H_{\mathrm{z}}$ and $\hat H_{\mathrm{ee}}$ commute with $\hat{\mathbf S}^2$ and therefore cannot couple $\ket{{}^1(TT)}$ to the high-spin manifolds; only the anisotropic $\hat H_{\mathrm{zfs}}$ breaks total-spin conservation and \emph{mixes} the manifolds, allowing population to leak out of the optically prepared singlet character into the dark high-spin states~\cite{Tayebjee2017,Collins2023}. The problem simplifies considerably for indistinguishable triplets, i.e., for aligned ZFS tensors at zero field: the symmetric $\ket{{}^1(TT)}$ decouples from the antisymmetric $\ket{{}^3(TT)}$ and couples only to a single accessible quintet state, reducing the spin dynamics to that of an effective two-level system, $\hat H_{\mathrm{TLS}} = -\tfrac{\Delta}{2}\hat\sigma_x - \tfrac{\varepsilon}{2}\hat\sigma_z$, in which the gap $\Delta$ is fixed by the zero-field splitting and the detuning $\varepsilon$ by the exchange coupling~\cite{Collins2023}.

\paragraph*{Singlet fission couplings.} The fission step itself is set by the coupling between the bright singlet $\ket{S_1}$ and the triplet pair. Because this coupling conserves total spin, $\ket{S_1}$ couples directly \emph{only} to the singlet-character pair $\ket{{}^1(TT)}$ and to no other spin component~\cite{Merrifield1971,Smyser2020,Casanova2018}. The microscopic origin of this matrix element remains debated, and two limiting mechanisms are usually invoked~\cite{Smith2010,Miyata2019}. In the \emph{direct} (one-step) mechanism, $\ket{S_1}$ and $\ket{{}^1(TT)}$ are connected by a two-electron exchange integral without real population of any intermediate. In the \emph{mediated} (two-step) mechanism, the coupling proceeds by superexchange through virtual, energetically higher-lying charge-transfer (charge-separated) states, which are only transiently populated; this pathway is the one most commonly held responsible for efficient fission in acenes~\cite{Casanova2018,Miyata2019,Chan2013}. In either case the primary step is spin-allowed and can proceed coherently and on sub-picosecond to picosecond timescales, as evidenced by real-time observation of the multiexcitonic intermediate, vibronic quantum beats, and conical-intersection-mediated dynamics in archetypal acenes~\cite{Chan2013,Bakulin2016,Musser2015,Miyata2019}. Subsequent ZFS-driven manifold mixing redistributes the singlet character among the triplet and quintet sublevels, and dephasing of the inter-triplet coherence finally yields two formally independent triplets~\cite{Smyser2020,Tayebjee2017,Scholes2015, Manian2023}. 

\subsection{Singlet fission in excitonic batteries}
\label{ss:sf-battery-model}

\noindent
In crystals, films, and aggregates the picture above becomes richer and requires careful control. Both the singlet and the correlated triplet pair can delocalise over several chromophores. Following the fission event, individual triplets can migrate by short-range Dexter hopping and other mechanisms, until the pair separates into freely diffusing triplets whose transport can span tens of nanometres~\cite{Miyata2019,Teichen2015,Smith2010}. This is simultaneously the central difficulty and the central opportunity. The difficulty is that disorder, the coupling to a phonon bath, and the multiplicity of nearly degenerate pair geometries make the spin and population dynamics far harder to isolate than in an engineered dimer~\cite{Miyata2019,Bakulin2016}. The opportunity is that extended media offer high triplet yields, long-range and potentially directional triplet transport, and---unlike isolated dimers---a macroscopic ensemble that can couple \emph{collectively} to a radiation field or to an acceptor layer~\cite{Teichen2015,Einzinger2019}. Exploiting this collective channel could be key to extend the storage time in excitonic quantum batteries, combining fast light absorption and scalable triplet generation.

\paragraph*{A model for collective fission and triplet harvesting.} To make this concrete, we propose a minimal model of $N$ molecular donor sites $n=1,\dots,N$ where fission can occur and from which triplets can transfer collectively to an acceptor layer, such as silicon or nanocrystal films. Following Sec.~\ref{ss:modelling-excitonic-qb}, we consider an organic cavity where the cavity serves as a charger and the organic molecules serve as battery, as shown in Eq.~\eqref{eq:total_hamiltonian}. For each molecule, i.e., site $i = \{1,\cdots,N\}$ in the donor layer, we retain a ground state $\ket{S_0}_i$, a bright singlet $\ket{S_1}_i$, and a degenerate triplet $\ket{T_1}_i$, as discussed in Eq.~\eqref{eq:local_basis}. We then use the singlet and triplet creation operators of Eqs.~\eqref{eq:singlet_creation} and~\eqref{eq:triplet_creation} to describe the different terms in the battery Hamiltonian $\hat{H}_B$, given by
\begin{equation}
    \label{eq:sf-battery}
    \hat{H}_B = \hat{H}_\mathrm{S}+\hat{H}_\mathrm{T} + \hat{H}_\mathrm{SF}.
\end{equation}

The first term represents the bright singlet degrees of freedom, including singlet local energies and hopping via dipole--dipole interactions
\begin{equation}
    \label{eq:singlet}
    \hat{H}_\mathrm{S}=\hbar\omega_S\sum_{i=1}^N \hat{\mathcal{S}}_i^\dagger\hat{\mathcal{S}}_i^\phdagger + \sum_{i<j} J^{(S)}_{ij} \left( \hat{\mathcal{S}}_i^\dagger\hat{\mathcal{S}}_j^\phdagger + h.c. \right),
\end{equation}
where $J_{ij}^{(S)}\propto r_{ij}^{-3}$ is the singlet hopping coupling strength, which depends on the distance $r_{ij}$ between two sites. The second term in Eq.\eqref{eq:sf-battery} represents the triplet degrees of freedom, here simplified to only account for one degenerate triplet state,
\begin{equation}
        \label{eq:triplet}
    \begin{split}
       \hat{H}_\mathrm{T}=&\hbar\omega_T\sum_{i=1}^N \hat{\mathcal{T}}_i^\dagger\hat{\mathcal{T}}_i^\phdagger + \sum_{i<j} J_{ij}^{(T)} \left( \hat{\mathcal{T}}_i^\dagger\hat{\mathcal{T}}_j^\phdagger + h.c. \right) \\
       &+\sum_{i<j} \chi_{ij} \hat{\mathcal{T}}_i^\dagger \hat{\mathcal{T}}_j^\dagger \hat{\mathcal{T}}_j^\phdagger\hat{\mathcal{T}}_i^\phdagger,
    \end{split}
\end{equation}
where $J_{ij}^{(T)} \propto \exp(r_{ij}/r_0)$ is the triplet hopping coupling strength, which is typically short-range and exponentially decaying in the distance $r_{ij}$ between sites, and where $\chi_{ij}$, also commonly short range, is the exchange interaction between triplets. Finally, the singlet fission term
\begin{equation}
    \label{eq:Hsf}
    \hat{H}_\mathrm{SF}=\sum_{i<j} \nu_{ij} \left(\hat{\mathcal{T}}^\dagger_i\hat{\mathcal{T}}^\dagger_j\hat{\mathcal{S}}_i^\phdagger+\hat{\mathcal{T}}^\dagger_i\hat{\mathcal{T}}^\dagger_j\hat{\mathcal{S}}_j^\phdagger+h.c.\right),
\end{equation}
where $\nu_{ij}$ is the effective singlet fission coupling strength. Similarly to Eq.~\eqref{eq:dicke}, the singlet manifold is coupled to the cavity mode via Dicke-like interactions
\begin{equation}
    \label{eq:singlet-cavity}
    \hat{H}_{BC}=g\sum_{i=1}^N \left(\hat{a}+\hat{a}^\dagger \right) \left( \hat{\mathcal{S}}_i^\phdagger+\hat{\mathcal{S}}_i^\dagger \right),
\end{equation}
with the cavity Hamiltonian $\hat{H}_C = \hbar\omega_c\hat{a}^\dagger\hat{a}$ being unchanged from Eq.~\eqref{eq:cavity-field}. 

As discussed in Sec.~\ref{ss:modelling-excitonic-qb}, the dynamics of the device is open and affected by loss channels such as the cavity leak at rate $\kappa$, and singlet fluorescence at rate $\gamma_S$. Here, we also consider a triplet-capture channel, representing triplet transfer from the molecules to some triplet-acceptor layer, e.g., crystalline rubrene ($\mathrm{C_{42} H_{28}}$)~\cite{Ma2012}. In particular, we propose to consider a collective triplet acceptor channel, given by 
\begin{equation}
    \label{eq:collective_triplet_acceptor}
    L_T^{(col)}=\sqrt{\Gamma}\sum_{i=1}^N \hat{\mathcal{T}}_i,
\end{equation}
where $\Gamma$ is the triplet capture rate. Such collective channel opens to capturing delocalised triplets, leading to the cooperative enhancement of the triplet capture rate, known as \textit{supertransfer}~\cite{Taylor2018, Kushwaha2025}, which scales linearly with the number $N_D$ of donor sites, i.e., the molecules hosting the triplets, and the number $N_A$ of acceptor sites.

\paragraph*{Harnessing collective effects.} This framework makes explicit the levers available for improving SF efficiency in an energy-storage context. First, engineering cooperativity through cavity coupling, molecular ordering, and concentration can enhance the collective absorption that charges the singlet population at a super-extensive rate~\cite{Ferraro2018,Quach2022}. Second, optimising singlet fission, as done in Ref.~\cite{Campaioli2024_SF}, toward the formation of delocalised triplets opens to the opportunity of engineering supertransfer. Together, these conditions are key to avoiding the local, i.e., not-scalable, triplet formation and capture rates of individual molecules and molecular dimers. The proposal that emerges is that of a molecular ensemble that absorbs superextensively, converts the absorbed energy into dark, weakly emissive triplet pairs that delocalise across the donor layer, with the opportunity for scalable triplet-capture by a stable acceptor layer. Realising this kind of device requires tuning cavity and molecules energy and couplings, minimising disorder and dephasing to prevent localisation, and identifying triplet-acceptor materials compatible with supertransfer and from which charges can be extracted on demand. 

\section{Charge-separated states}
\label{s:charge-separated}

A charge-separated state is formed when a photoexcited electron and its accompanying hole come to reside on physically distinct molecular species, rather than remaining bound as a Frenkel exciton on a single chromophore. 
This spatial separation removes the strong electron-hole Coulomb attraction that drives radiative and non-radiative recombination in a localised exciton, and it is precisely this property that makes charge-separated states attractive both for organic photovoltaics~\cite{Fukuzumi2014} and, more recently~\cite{Hymas2026}, for excitonic quantum batteries: a sufficiently deep type-II heterojunction can trap energy~\cite{Lo_AM2011a} in a configuration from which spontaneous return to the ground state is kinetically suppressed, while simultaneously presenting that energy in a form, separated charge, that is already compatible with extraction as an electric current.

\begin{figure*}
    \includegraphics[width=\linewidth]{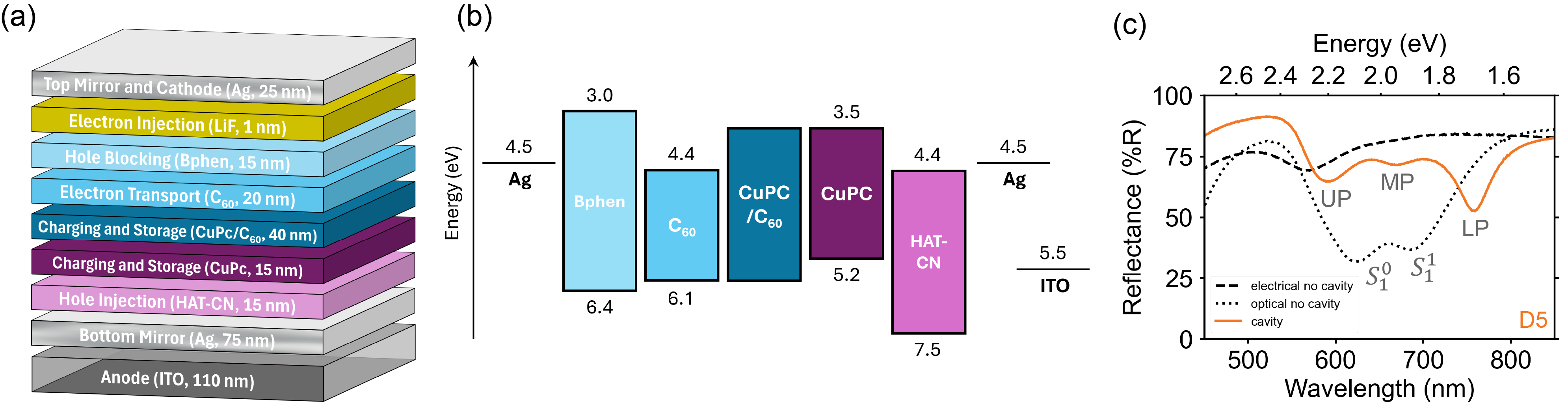}
    \caption{\textbf{Full-cyle quantum battery device design.}---(a) Schematic of the layered structure of the quantum battery, describing the function and composition of each component. 
    Ultrafast pump and probe laser pulses are used to charge and measure the superextensive charging of the device. 
    To characterise the system outside the cavity, electrical control devices are fabricated by removing the bottom mirror, thereby eliminating the cavity resonance and its related effects while keeping electrical connection intact.
    Optical control devices are fabricated by removing the top mirror and retaining the base mirror, to allow for optical reflectance measurements. 
    (b) Work functions and HOMO/LUMO energy levels of each layer in the quantum battery, defining the energy gradient for charge separation.
    (c) Steady-state reflectance spectra at 8$^\circ$ AOI for the quantum battery (orange), and the electrical control (black dashed) and optical control (black dotted) devices, which have no cavity layer. 
    The control is characterised by singlet states $S^0_1$ and $S^1_1$. 
    In the quantum battery, these singlet states are hybridised with the photonic cavity mode to give rise to polaritons, as characterised by the upper (UP), middle (MP), and lower (LP) polariton states.
    Adapted with permission from \citet{Hymas2026}.}
    \label{fig:hymas}
\end{figure*}

Our full-cycle device, reported in  \citet{Hymas2026} and introduced in Sec.~\ref{s:molecular_triplets}, realises this mechanism directly. 
The CuPc layer responsible for superabsorption and triplet storage \textit{via} intersystem crossing is interfaced with C$_{60}$ acceptor, forming a well-characterised type-II donor-acceptor heterojunction~\cite{Uchida_APL2004,Dutton_JPCC2012}, as shown in Fig.~\ref{fig:hymas}a. 
Both the photoexcited singlet and the metastable triplet excitons generated in CuPc diffuse toward this interface, where they can form an interfacial charge-transfer state before separating into a free electron in C$_{60}$ and a free hole in CuPc. 
Crucially, this charge-separation step is supported by an engineered energy gradient spanning the full device stack, outlined in Fig.~\ref{fig:hymas}b: 
the HAT-CN, BPhen, and LiF layers surrounding the donor-acceptor junction are chosen not for their optical properties, but for their HOMO/LUMO alignment, which creates an energy-level gradient in the optical microcavity that biases charge transport in the forward direction, while kinetically blocking the reverse flow of carriers that would otherwise recombine. 
The result is a device in which the molecular triplet manifold discussed in Sec.~\ref{s:molecular_triplets} functions simultaneously as a longer-lived energy storage reservoir and as the source population for charge separation, linking the two mechanisms reviewed in this chapter into a single operational pathway from absorbed photon to extracted charge.

This device architecture shows clear strong coupling in the cavity, identified as upper (UP), middle (LP) and lower (LP) polariton reflectance signatures in the orange trace of Fig.~\ref{fig:hymas}c, compared against the reflectance spectra of electrical (black dashed trace) and optical (black dotted trace) control devices which show no signs of cavity-mediated collective effects.
The bare exciton absorption peaks of CuPc are identified in the optical control device, labelled $S^0_1$ and $S^1_1$.

This architecture is significant for excitonic quantum batteries beyond its role as a proof of concept, because it is the first instance in this exciton platform where the collective light-matter coupling responsible for superabsorption demonstrably survives the full charging-storage-extraction cycle and leaves a signature in an electrical, rather than purely optical or spectroscopic, observable.

\begin{figure}
    \centering
    \includegraphics[width=0.75\linewidth]{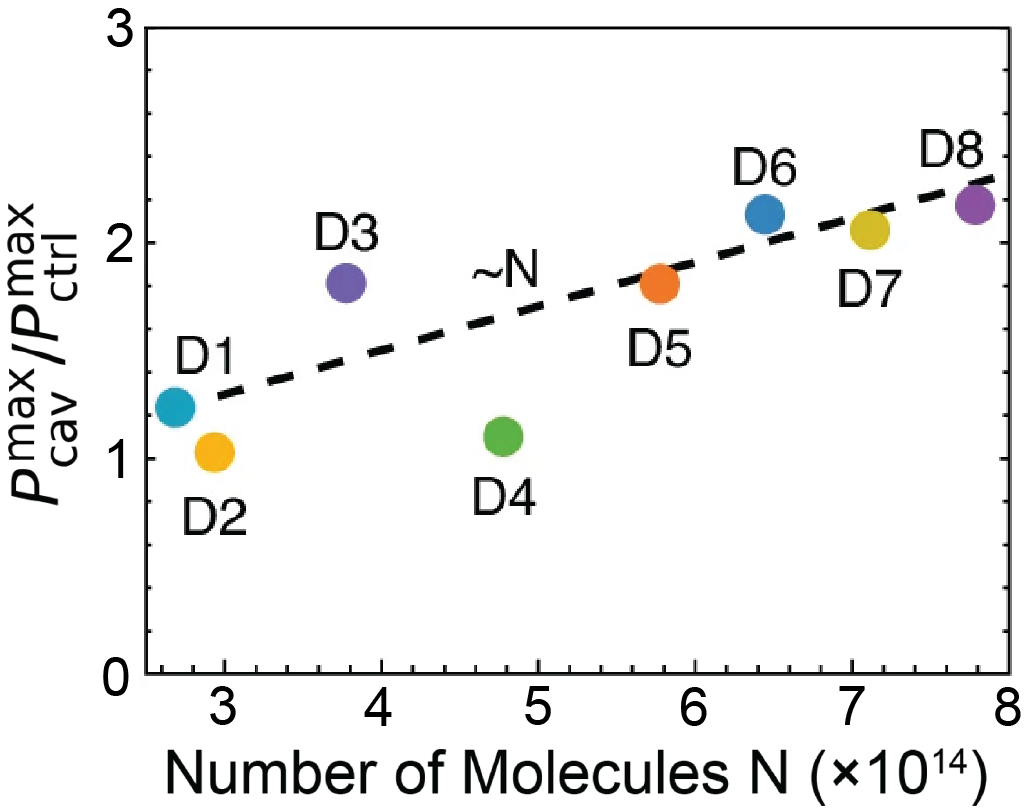}
    \caption{\textbf{Superextensive scaling of charging power in a model quantum battery.---}
    Systematic increases in the ratio $P^\mathrm{max}_\mathrm{cav} / P^\mathrm{max}_\mathrm{ctrl}$ with $N$, determined from photocurrent-voltage measurements, indicate superextensive scaling of discharging power, consistent with a collective extraction mechanism enabled by strong coupling.
    Adapted with permission from~\citet{Hymas2026}.}
    \label{fig:hymas_PvN}
\end{figure}

We observed that the ratio of peak discharging power between the cavity device and its ``no-cavity" control grows linearly with the number of absorbers $N$, which we attributed to a polariton-dressed open-circuit voltage that scales as $\sqrt{N}$, yielding an overall discharging power 
$P_\mathrm{cav}^\mathrm{max} \propto N^2$.
This relationship in shown in Fig.~\ref{fig:hymas_PvN}, where the series of eight devices with increasing number of absorbers $N$ show a superlinear scaling in discharging power.
This is a markedly different statement from the superextensive \textit{charging} power discussed throughout this chapter: it demonstrates that collective enhancement can persist all the way to the steady-state electrical output of the device, under continuous, incoherent illumination, and at room temperature. 

It is worth emphasising, however, our device published in \citet{Hymas2026} is not, strictly speaking, a battery in the sense implied by the rest of this chapter: energy is continuously photogenerated and extracted under steady illumination, rather than charged, stored, and subsequently discharged on demand. 
The charge-transport layers responsible for extraction are permanently connected to the storage layer, so the device behaves as a cavity-enhanced \textit{photodiode}~\cite{Eizner_AP2018} with a metastable intermediate, rather than as a battery with a controllable charge-store-discharge cycle. 

Realising a battery in the strict sense based on this architecture will require a device in which superabsorption persists undisturbed by the incoherent processes of charge separation and accumulation, while the resulting separated charge is held in place and its electrical discharge is gated and triggered only on demand.
Nevertheless, the demonstration that collective effects can enhance the rate of charge extraction itself, \textit{surviving the inherently incoherent nature of charge separation and transport}, is an important and previously unobserved result, and one we hope will spur further experimental and theoretical efforts toward realising a genuine charge-separated quantum battery.

\section{Conclusions and perspectives}
\label{s:conclusions}

\noindent
In this chapter we have examined three design principles to extend the energy storage lifetime of excitonic quantum batteries, each addressing the same underlying tension: the collective coupling that enables superextensive charging through superabsorption can also drive rapid, collective discharge~\cite{Quach2022, Campaioli2024}. 
The common remedy is to charge through a bright manifold of photoactive states and store energy on a dark manifold, by controlling the coupling between the two. Transferring energy to long-lived localised molecular triplets via intersystem crossing or polariton--triplet coupling can yield significant storage lifetime extensions, as shown in Ref.~\cite{Tibben_PRX2025}. However, this approach faces three limitations. First, triplet lifetimes are often limited to the microsecond range in most materials that combine fast intersystem crossing with cavity coupling. Second, the singlet-to-triplet conversion dissipates energy set by the singlet--triplet gap, which is lost as heat. Third, triplet formation and transfer in this mechanism are limited to individual chromophores and thus cannot scale with the number $N$ of molecules. 

Singlet fission partially resolves two of these issues. In this spin-allowed process, one bright singlet is converted into two dark triplets on ultrafast timescales, with each triplet carrying roughly half the singlet energy, so that the conversion is approximately energy-conserving, outcompeting single-molecule radiative loss rates. More importantly, when singlet fission is lifted from molecular dimers to a collectively coupled ensemble, the triplet pair can be generated in a delocalised state~\cite{Campaioli2024_SF}, opening a route to scalable triplet harvesting via supertransfer to an acceptor layer~\cite{Taylor2018, Kushwaha2025}, provided that the delocalised triplets, or the charges they produce, are captured rapidly enough to outpace recombination.

Recent groundbreaking results by Soto~\etal~\cite{Soto2026} suggest that a departure from the traditional organic fluorescent molecules considered so far 
could open to triplet state lifetimes that exceed 10 hours. Indeed, Soto~\etal~\cite{Soto2026}, reported triplet states of carbon-germanium germylenes  with 14-h half-life at room temperature, marking an improvement of up to 10 orders of magnitude over the typical microsecond lifetime of molecular triplet states in organic dyes. However, the strategy is so far limited to one ligand-stabilized system, and its generality across other tetrylenes, robustness under ambient conditions, and scalability remain open questions for future work.

Finally, charge-separated states offer another route to push storage lifetimes beyond the microsecond scale and into a regime meaningful for practical optoelectronic applications. These are the longest-lived electronic excitations in molecular systems, with promising reports of 1.2 second lifetimes in organic crystalline nanoparticles~\cite{Cai2026}, 2 hour lifetime for charge-separated states in covalently-linked donor-acceptor molecules~\cite{Fukuzumi_JOTACS2004a} to month-long electron lifetime in disordered organic solid-state films~\cite{Yamanaka2023}. These suggest storage times relevant to practical optoelectronic devices~\cite{Fukuzumi2014, tang2023manipulating}, although their use as a storage register for excitonic quantum batteries is only beginning to be explored, and much work remains to be done in this direction.

Despite progress and promising outlooks, several challenges separate excitonic quantum batteries from practical application. The most immediate is energy extraction: a battery is defined not by the energy it stores but by the work it can return, and nearly all experiments to date report stored energy or excited-state populations inferred from transient spectroscopy, rather than the extractable work, or ergotropy, that ultimately characterises performance~\cite{Allahverdyan2004}. Making extraction operational will require coupling the storage layer to an external circuit, for which the charge-transport layers already used to read out charge-separated states offer a natural starting point~\cite{Hymas2026}. 

On the theory side, two idealisations underlying the models reviewed here should eventually be relaxed. The first is the reduction of each molecule to a few electronic levels, which neglects the structured vibrational environment that dresses real chromophores. Vibronic coupling and non-Markovian bath memory shape both the population and the spin dynamics, and capturing them faithfully requires the explicit treatment of system--bath correlations rather than a memoryless reservoir~\cite{Strathearn_NC2018, Fux_JCP2024}. The second is the Dicke description itself, which assumes identical emitters coupled uniformly to a single cavity mode and thus discards the spatially varying, intrinsically multi-mode character of realistic light--matter coupling; establishing the regime of validity of this approximation, and the robustness of the conclusions beyond it, remains an open task~\cite{Kockum2019}. 
Finally, there are the practical challenges of fabrication, where realising the spatial separation of charging and storage layers, and ultimately the controlled placement of emitters at field antinodes, calls for a level of structural precision beyond that of the disordered, solution-processed films used so far. None of these obstacles appears fundamental, and each defines a concrete direction in which the platform can be improved, from device architecture and material selection to the theoretical tools used to design them, defining the roadmap for the development of excitonic quantum batteries.

Beyond organic microcavities, the strategies to extend energy storage lifetimes that we have reviewed here can be extended and generalised to other platforms, such as trapped ions, neutral atoms, and superconducting circuits. Every approach discussed in this chapter aims to shape the relaxation of an open many-body system so that a chosen manifold is reached quickly yet decays slowly. Engineering metastability in driven--dissipative quantum systems is itself an active area of investigation~\cite{Macieszczak2016}. Recent progress on anomalous relaxation offers a constructive path based on engineering the spectrum of the Liouville superoperator $\mathcal{L}$ generating the open dynamics of Eq.~\eqref{eq:lindblad} to accelerate or slow-down the approach to equilibrium~\cite{Carollo2021, Yin2024, Teza2026, Beato2026}.  Porting this metastability-engineering toolbox across platforms, and feeding the lessons learned back into organic microcavities, plasmonic materials, and other room-temperature settings, could accelerate progress on all these fronts.


\begin{acknowledgments}
\noindent
D.G. and D.T. acknowledge support from the Australian Research Council DP230101764.
\end{acknowledgments}

\bibliography{main.bbl}

\vfill
\clearpage

\end{document}